\documentclass[10pt,prx,twocolumn,amsmath,amssymb,floatfix,notitlepage]{revtex4-1}
\usepackage[T1]{fontenc}
\usepackage[utf8]{inputenc}
\usepackage{lmodern}
\usepackage{microtype,bm,bbm}
\usepackage{graphicx,booktabs}
\usepackage{times}
\usepackage[usenames,dvipsnames]{xcolor}

\graphicspath{{figs/}}

\usepackage{hyperref}
\hypersetup{
  colorlinks,
  allcolors=NavyBlue,
  linktoc=all
}

\usepackage[capitalize]{cleveref}
\crefformat{section}{Sec.~#2#1#3}

\newcommand{\D}{\mathcal D}

\renewcommand*\O{\mathcal{O}} 
\newcommand*\dagg{^{\dagger}}
\newcommand*\eps{\varepsilon}
\newcommand{\id}{\mathbbm{1}}
\newcommand*\mat[1]{\begin{pmatrix}#1\end{pmatrix}} 
\newcommand*\matr[1]{\mathsf{#1}}
\renewcommand*\vec[1]{\mathbf{#1}}

\newcommand*\ket[1]{|#1\rangle}
\newcommand*\bra[1]{\langle #1|}

\DeclareMathOperator{\erf}{erf}

\DeclareMathOperator{\tr}{tr}
\let\Re\relax
\let\Im\relax
\DeclareMathOperator{\Re}{Re}
\DeclareMathOperator{\Im}{Im}

\begin{document}
\title{Nondestructive photon counting in waveguide QED}
\author{Daniel Malz}
\author{J.\ Ignacio Cirac}
\affiliation{Max Planck Institute for Quantum Optics, Hans-Kopfermann-Stra{\ss}e 1, D-85748 Garching, Germany}
\affiliation{Munich Center for Quantum Science and Technology, Schellingstra{\ss}e 4, D-80799 München, Germany}
\date{\today}

\begin{abstract}
  Number-resolving single-photon detectors represent a key technology for a host of quantum optics protocols, but despite significant efforts, 
  state-of-the-art devices are limited to few photons.
  In contrast, state-dependent atom counting in arrays can be done with extremely high fidelity up to hundreds of atoms.
  We show that in waveguide QED, the problem of photon counting can be reduced to atom counting, by entangling the photonic state with an atomic array in the collective number basis.
  This is possible as the incoming photons couple to collective atomic states
  and can be achieved by engineering a second decay channel of an excited atom to a metastable state.
  Our scheme is robust to disorder and finite Purcell factors, and its fidelity increases with atom number.
  Analyzing the state of the re-emitted photons, we further show that if the initial atomic state is a symmetric Dicke state,
  dissipation engineering can be used to implement a nondestructive photon-number measurement,
  in which the incident state is scattered into the waveguide unchanged.
  Our results generalize to related platforms, including superconducting qubits.
\end{abstract}
\maketitle

\section{Introduction}
Single-photon detectors have a long history~\cite{Morton1949}, with a plethora of technologies available~\cite{Hadfield2009}.
Applications in quantum optics, such as quantum state preparation, quantum metrology~\cite{Giovannetti2011}, entanglement distribution~\cite{Gisin2002}, and quantum computing~\cite{Knill2001,Kok2007}
have placed a renewed focus on single-photon detectors capable of resolving the number of incoming photons.
Perhaps most promising are superconducting transition-edge sensors,
which have been demonstrated to achieve a (per-photon) detection efficiency (percentage of detected photons) of $\eta\approx95\%$~\cite{Lita2008} and to distinguish up to seven photons,
with a negligible dark count rate (clicks in the absence of incoming photons).
They are based on the principle that near the critical temperature,
the resistance of a superconductor is very sensitive to temperature changes, down to the level of single-photon energies.
While very impressive, the device is limited to optical photons, destroys the photonic state,
and is difficult to scale.

One strategy that in principle also allows for nondestructive measurements is based on quantum memory. 
An itinerant photon may be caught by an atom in a cavity~\cite{Specht2011} or by a cavity with tunable coupling~\cite{Yin2013}, which however requires time-dependent tuning of the atom--cavity or cavity--waveguide coupling in accordance with the photon wavepacket shape.
More generally, electromagnetically-induced transparency (EIT) can be used to slow down a light pulse such that it fits within an atomic cloud~\cite{Fleischhauer2000,Fleischhauer2005}.
This allows the storage of arbitrary photon pulses within a length given by system parameters. 
The number of polaritons can in principle be read out by employing a cycling transition~\cite{Imamoglu2002,James2002}.
However, this scheme also requires knowledge about the arrival of the photon wavepacket.
Furthermore, the combination of quantum memory and quantum nondemolition detection of one or few excitations in a 3D cloud of atoms makes this very challenging.
EIT in a waveguide QED setting is one strategy to alleviate the imaging problem (see, e.g., Ref.~\cite{Caneva2015}).

Here, we instead design a detector that does not require knowledge about the shape or arrival time of incoming wavepackets and therefore does not require implementation of a time-dependent Hamiltonian.
This can be achieved either through continuous measurement, for example through dispersive coupling~\cite{Chen2011a,Govia2012,Poudel2012,Kyriienko2016,Oelsner2017,Schondorf2018},
or if the detector permanently changes its state and is read out later,
as in impedance-matched $\Lambda$-systems~\cite{Pinotsi2008,Romero2009,Romero2009a,Peropadre2011,Koshino2013,Inomata2016,Bechler2018}.
The former class of detectors suffer from measurement backaction on the photonic state~\cite{Helmer2009,Royer2018},
whereas the latter does not produce `clicks' to indicate detection event, and therefore does not localize photons in the waveguide, which avoids measurement backaction.
The advantage of this strategy is that no knowledge about the photon wavepacket is needed, and that photons do not have to be destroyed to be detected,
which opens the way toward nondestructive photon-number measurements.

In recent years, quantum emitters coupled to waveguides have emerged as a powerful experimental paradigm~\cite{Chang2018,Turschmann2019}. 
Examples of such systems include cold atoms levitated near optical fibres~\cite{Vetsch2010,Douglas2015} or photonic crystal waveguides~\cite{Goban2015}, but also solid-state realizations such as quantum dots~\cite{Akimov2007,Lodahl2015}, superconducting qubits~\cite{VanLoo2013,Sundaresan2019},
or nitrogen-vacancy centres~\cite{Huck2011,Sipahigil2016,Evans2018}. 
Due to the strong confinement of light, even a single emitter can have a profound effect on light propagation,
and many emitters show remarkable collective effects~\cite{Solano2017}.
At the same time, impressive experimental breakthroughs have enabled the control~\cite{Schlosser2001} and, in particular, readout of neutral atoms~\footnote{In an array of 160 atoms, a recent experiment achieved $99.94\%$ readout fidelity, albeit at the expense of a constant atom loss rate that dominates this theoretical fidelity~\cite{Wu2019}. Atoms trapped in tweezer arrays can circumvent this problem, offering similar fidelity without losing the atoms~\cite{Covey2019}.}, superconducting qubits~\cite{Jeffrey2014,Walter2017}, and trapped ions~\cite{Harty2014}.
Thus, quantum emitters coupled to waveguides appear to be an ideal platform to produce~\cite{Goban2015,Paulisch2018,Paulisch2019,Perarnau-Llobet2019} and control~\cite{Zheng2013,Zhang2019} quantum light.
Here we show that they also allow for detection of quantum light.

\begin{figure}[t]
  \centering
  \includegraphics[width=\linewidth]{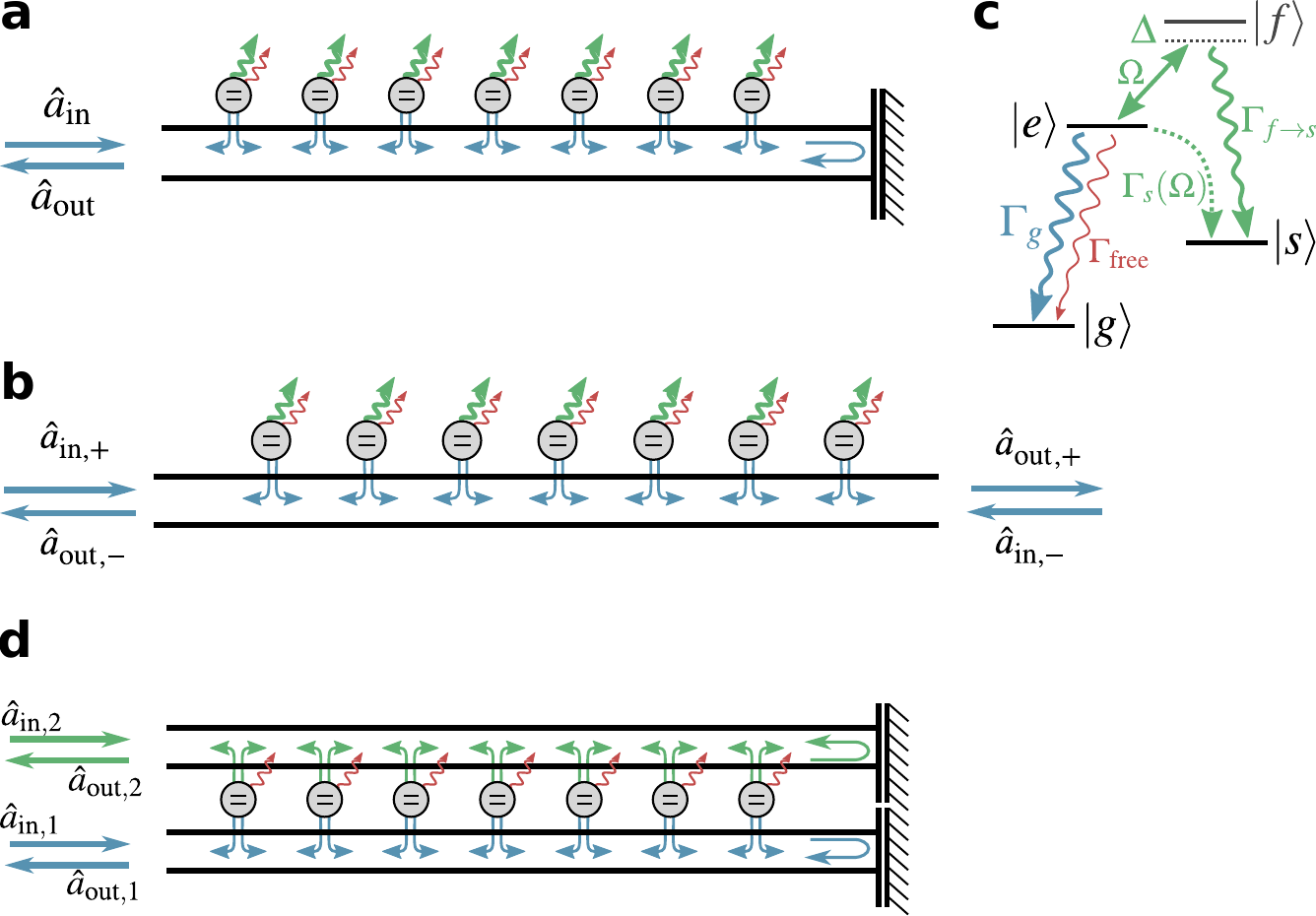}
  \caption{
  	\textbf{Sketch of the proposed setups.}
  	We consider arrays of quantum emitters coupled to (\textbf{a}) a semi-infinite waveguide terminated by a mirror at $x=0$,
  	(\textbf{b}) an infinite waveguide, and (\textbf{d}) to two waveguides.
  	The decay $e\to g$ couples to the waveguide (blue, $\Gamma_g$), but also to free-space modes (red, $\Gamma_{\mathrm{free}}$).
  	Key ingredient is a tunable engineered decay from the excited $\ket e$ to a metastable state $\ket s$ (green, $\Gamma_s(\Omega)$),
  	implemented for example as a Raman transition as shown in \textbf{c}.
  	The photons may be emitted either into free space (\textbf{a, b}, \cref{sec:destructive}) or back into the waveguide 
  	for a nondestructive measurement (\textbf{d}, \cref{sec:QND}).
  }
  \label{fig:sketch}
\end{figure}

The key idea here is to engineer atoms such that for each incident photon in the waveguide,
exactly one atom changes its internal state, 
such that a subsequent measurement of the atomic state yields the number of photons in the scattered wavepacket.
We do this by identifying conditions such that all photons are absorbed in one atomic transition ($g\to e$, blue in \cref{fig:sketch})
and dissipated in a different one ($e\to s$, green).
This way, the atomic array keeps a memory of the number of scattered photons~\cite{Pinotsi2008,Romero2009,Romero2009a,Peropadre2011,Koshino2013,Inomata2016}.

In order to identify such conditions, we study a realistic model for photon absorption by atomic arrays in the presence of free-space decay (red) and disorder.
In view of experimental realizations, we provide recipes for engineering an additional decay channel,
and pay particular attention to spatial disorder, as this fundamentally modifies the eigenmodes of the system.
Surprisingly, we find that it is possible to engineer dissipation to achieve full absorption independent of disorder.
As a consequence of collective enhancement in the atom-waveguide coupling, detection efficiency and bandwidth grow with atom number,
such that even with moderate Purcell factor (ratio of waveguide to free-space decay rate),
detection efficiencies $\eta>99\%$ can be reached for intermediate numbers of atoms ($N>20$).
We also consider emitters coupled to chiral waveguides, which have distinct advantages due to the natural suppression of backscattering.
Our detection scheme is scalable, as the number of atoms required to detect a certain number of photons with a fixed error grows only polynomially.

In a second part, we study a natural modification, in which the dissipated photons are emitted back into the waveguide (\emph{cf.} \cref{fig:sketch}d).
If the the resulting output wavepacket coincides with the input wavepacket, this realizes a quantum nondemolition (QND) measurement~\footnote{We use the terminology adopted by most authors~\cite{Thorne1978,Unruh1978,Braginsky1980,Caves1980,Nogues1999,Raimond2001,Guerlin2007,Wiseman2009,Johnson2010}, but note that there is some controversy whether this is an appropriate name~\cite{Monroe2011}, and sometimes \emph{nondestructive} is used instead~\cite{Reiserer2013}, but this term lacks the precise definition given by, \emph{e.g.} Wiseman and Milburn~\cite{Wiseman2009}.
}.
In order to achieve high fidelities, this requires collective enhancement of both decay channels,
which requires the atomic array to be prepared in a symmetric Dicke state between ground states.
We show that such states can be prepared through coherent interaction of two arrays and subsequent conditioning.
Ultimately, this can reach a scaling where for a given error probability, the number of atoms that can be prepared in this way scales exponentially with Purcell factor $N\sim\exp(P)$, which we verify with numerical simulations.

The rest of this paper is structured as follows.
In \cref{sec:model} we present the scattering theory for atomic arrays on waveguides.
Building on the principle of coherent perfect absorption, we show explicitly that perfect absorption can be obtained by tuning only dissipation, even in the presence of arbitrary disorder.
We apply this principle to establish the feasibility of number-resolving detection by absorption in a mirror geometry (\cref{sec:mirror}) and in an infinite waveguide (\cref{sec:infinite}).
For completeness, we also discuss how dissipation can be used in a chiral waveguide to obtain a number-resolving detector (\cref{sec:chiral}).
We extend these concepts to a QND measurement in \cref{sec:QND}.
In \cref{sec:experiment}, we discuss the experimental implementations of our proposal, including engineered dissipation, atomic species, readout, dector reset, 
the effect of non-idealities, and bandwidth. We conclude in \cref{sec:conclusion}.

\section{Absorption in atomic arrays}\label{sec:model}
Below, we first formulate a generic input-output theory for photons scattering off atomic arrays (\cref{sec:scattering}), following previous work~\cite{LeKien2008,LeKien2014,Caneva2015}.
This allows us to generically identify the conditions under which all photons are absorbed in one transition and emitted in the other (\cref{sec:absorption}).

\subsection{Scattering theory}\label{sec:scattering}
A generic Hamiltonian describing the system-waveguide interaction is given through~\cite{LeKien2008,LeKien2014,Caneva2015}
\begin{equation}
  \begin{aligned}
  H=&\int_0^\infty\frac{dk}{2\pi}\sum_{\nu}\sum_{i=1}^2(\omega_{k,i}-\omega_0)a_{\nu,k,i}\dagg a_{\nu,k,i}\\
  {}-{}&\sum_{n,\nu}\left(g_{\nu,k,1}^{(n)} \sigma_{eg}^{(n)}a_{\nu,k,1}+g_{\nu,k,2}^{(n)} \sigma_{es}^{(n)}a_{\nu,k,2} +\text{H.c.}\right)
  \end{aligned}
  \label{eq:generic_hamiltonian}
\end{equation}
In \cref{eq:generic_hamiltonian}, the label $\nu$ runs over different sets of waveguide modes,
the index $i$ denotes whether the waveguide mode couples to the $g\leftrightarrow e$ transition ($i=1$) or the $s\leftrightarrow e$ transition ($i=2$),
the index $n$ runs of the $N$ atoms,
the individual waveguide modes are labelled by their wavevector $k$,
and the coupling of them to the atoms is given by the rate $g_{\nu,k,i}^{(n)}$. 

This Hamiltonian describes a situation where the two transitions are coupled to different fields, which can either be waveguide modes or free-space modes.
In an infinite waveguide (\cref{fig:sketch}b),
$g_{\nu,k,1}^{(n)}=\sqrt{2c\Gamma_{g}}\exp(ik\nu x_n)$ and $\nu\in\pm$, corresponding to left- and right-moving modes,
whereas for the semi-infinite waveguide (\cref{fig:sketch}a),
there is only one set of waveguide modes with coupling $g_{k,1}^{(n)}=\sqrt{c\Gamma_{g}}\sin(kx_n)$.
In these expressions, $\Gamma_{g}$ is the decay rate to $\ket g$ of an individual atom into the waveguide, and $\omega_k=ck$.
Integrating out the bath modes yields quantum Langevin equations for the spin operators (see \cref{app:langevin_equations}),
\begin{equation}
  \begin{aligned}
   	\dot \sigma_{ge}^{(n)}=(\sigma_{gg}^{(n)}-\sigma_{ee}^{(n)})&[\matr L_{n\nu,1}\vec a_{\mathrm{in},\nu,1}-i\matr H_{\mathrm{eff},1,nm}\sigma_{ge}^{(m)}]\\
   	{}+\sigma_{gs}^{(n)}&[\matr L_{n\nu,2}\vec a_{\mathrm{in},\nu,2}-i\matr H_{\mathrm{eff},2,nm}\sigma_{se}^{(m)}],
  \end{aligned}
  \label{eq:langevin}
\end{equation}
where the sum over input fields $\nu$ and atoms $m$ is implied.
The non-Hermitian Hamiltonians $\matr H_{\mathrm{eff},1}$ ($\matr H_{\mathrm{eff},2}$) describes both coherent interaction of the quantum emitters and decay into the waveguide induced by the coupling of the $g\leftrightarrow e$ ($s\leftrightarrow e$) transition to the waveguide, 
and $\vec a_{\mathrm{in},\nu,i}$ are waveguide fields coupling to these transitions. 
This coupling is given by the generically non-square matrix $\matr L_i$.
If coupled via an infinite waveguide, the $i^{\mathrm{th}}$ transition couples to two input fields, $\vec a_{\mathrm{in},i}=(a_{\mathrm{in,}+,i},a_{\mathrm{in},-,i})$, corresponding to right- and left-moving photons, such that $\matr L_i$ is a $N\times2$ matrix.
This is in contrast to a semi-infinite waveguide, which only has on input field, such that $\matr L_i$ is a vector.
Independent free-space decay at rate $\Gamma_{\mathrm{free},i}$, can also be captured by adding a term $\matr H_{\mathrm{free},i}=-i\Gamma_{\mathrm{free},i}\id/2$
$\matr L_3=\sqrt{\Gamma_{\mathrm{free}}}\id$, and introducing $N$ independent noise operators $\vec a_{\mathrm{in, free},i}$.
Note that the assumption of independent decay into free-space does not necessarily hold and depends on how closely spaced the quantum emitters along the waveguide are. 
Such a situation may lead to a reduction in free-space decay~\cite{Asenjo-Garcia2017}, which would benefit the detector proposed here.
Multiple baths coupling to the same transition appear as several terms taking either the form of the first or the second line on the right-hand side.

The Langevin equation for $\sigma_{se}^{(n)}$ can be obtained by exchanging both $1\leftrightarrow2$ and $g\leftrightarrow s$ everywhere in \cref{eq:langevin}.
The explicit form of $\matr H_{\mathrm{eff}}$ for the mirror geometry and the infinite waveguide are given in \cref{eq:mirror_Hamiltonian,eq:infinite_Heff} below.

If there are only few excitations compared to the number of atoms, one can linearize the Langevin equation using a Holstein-Primakoff transformation that sends $\sigma_{ge}\to b$~\cite{gardiner2004quantum,Chang2012,Caneva2015,Chang2018}.
In this approximation, $\sigma_{gs}^{(n)}\sigma_{se}^{(m)}\to\delta_{mn}\sigma_{ge}^{(n)}$, such that only the diagonal terms in the second line of \cref{eq:langevin} survive $\matr H_{\mathrm{eff},2}\to-i\Gamma_s\id/2$ (green in \cref{fig:sketch}), independent of whether the decay $e\to s$ corresponds to guided or non-guided modes.
Combining this with free-space decay (shown in red in \cref{fig:sketch}), we obtain uniform incoherent decay
at rate $\Gamma'=\Gamma_{\mathrm{free},1}+\Gamma_s$.
We thus arrive at
\begin{equation}
  	\dot{\vec b} = (-i\matr H_{\mathrm{eff},1}-\Gamma'/2)\vec b+\matr L_{\nu,1}\vec a_{\mathrm{in},\nu,1}.
  	\label{eq:generic_langevin_equations}
\end{equation}
Here, we have neglected all input fields except the ones pertaining to the waveguide, which is valid if they are in vacuum.
In order to describe destructive photon measurements, it is then sufficient to show that all incoming photons are absorbed and re-emitted into the bath modes coupling to the $s\leftrightarrow e$ transition.
Strictly speaking, decay to $\ket s$ eliminates the atom from the dynamics, but this effect is neglected in the linearization.

\subsection{Complete absorption}\label{sec:absorption}
We now examine in general how tuning $\Gamma'$ in \cref{eq:generic_langevin_equations} can lead to complete absorption of photons by an atomic array.
For the rest of this section, we drop the indices from $\matr H_{\mathrm{eff}}$ and $\matr L$, for sake of generality and simplicity.
In order to find the linear scattering properties of the array,
we use the input-output equations that relate the output field operators to the input fields
$\vec a_{\mathrm{out}}(t) = \vec a_{\mathrm{in}}(t)-\matr L\dagg \vec b(t)$~\cite{gardiner2004quantum}.
Solving \cref{eq:generic_langevin_equations}, we obtain the scattering matrix
\begin{equation}
  \begin{aligned}
  	\vec a_{\mathrm{out}}(\omega)
  	&=\left\{ \matr 1-\matr L\dagg\left[(\Gamma'/2-i\omega)\mathbbm1+i\matr H_{\mathrm{eff}}\right]^{-1}\matr L \right\}\vec a_{\mathrm{in}}(\omega)\\
  	&\equiv \matr S(\omega)\vec a_{\mathrm{in}}(\omega).
  \end{aligned}
  \label{eq:generic_scattering_matrix}
\end{equation}

A detector that counts the number of photons in a specific input port (say, $\vec a_{\mathrm{in},+}$) 
needs to absorb all of them and dissipate them via the transition to $\ket s$.
This is captured by our key figure of merit, the \emph{detection efficiency} $\eta$,
which is the product of the probability that a photon is not reflected
$p_{\mathrm{abs}}=1-\sum_{\nu\neq1}|\matr S_{\nu1}(\omega)|^2$, and the probability that it is dissipated via the engineered channel (rate $\Gamma_s$)
rather than into free space, $\eta=p_{\mathrm{abs}}\Gamma_{s}/(\Gamma_{s}+\Gamma_{\mathrm{free}})$.
Thus, a high fidelity requires $\Gamma_{s}\gg\Gamma_{\mathrm{free}}$ and $p_{\mathrm{abs}}\approx1$.
In waveguide QED this regime can be reached through the collective enhancement of the emitter--waveguide coupling as compared to the free-space decay.

Unity absorption is attained if one of the eigenvalues of the scattering matrix $\matr S(\omega)$ is zero.
This corresponds to a pole of the inverse scattering matrix
$\matr S^{-1}=[\matr 1+\matr L\dagg(-i\omega+i\matr H_{\mathrm{eff}}\dagg+\Gamma'/2)^{-1}\matr L]$.
A pole of $\matr S^{-1}$ arises whenever $\omega$ coincides with an eigenvalue of $\matr H_{\mathrm{eff}}\dagg-i\Gamma'/2$,
which implies that the scattering matrix $\matr S$ has a zero if $\omega-i\Gamma'/2$ coincides with an eigenvalue of $\matr H_{\mathrm{eff}}$.
Thus, tuning $\omega$ and $\Gamma'$, this can always be achieved, for any $\matr H_{\mathrm{eff}}$.
Absorption based on this principle has been observed in a variety of systems~\cite{Yan1989,Kishino1991,Cai2000},
and been termed coherent perfect absorption~\cite{Chong2010,Baranov2017}.
Note that to reach this conclusion we did not have to assume anything about the form of $\matr H_{\mathrm{eff}}$ (apart from linearity),
which is the reason it works for arbitrary disordered systems.
If these conditions are fulfilled, there exists an eigenvector $\vec e_0$, such that $\matr S(\omega_0)\vec e_0=0$.
Generically, $\vec e_0$ describes a linear combination of various input fields of the system,
which implies that coherent perfect absorption is only a sufficient condition for unity absorption property when there is only one input field, such as in the mirror geometry.
Nevertheless, in the infinite waveguide efficient detection is still attained for large atom numbers.

\begin{figure*}[t]
  \centering
  \includegraphics[width=\linewidth]{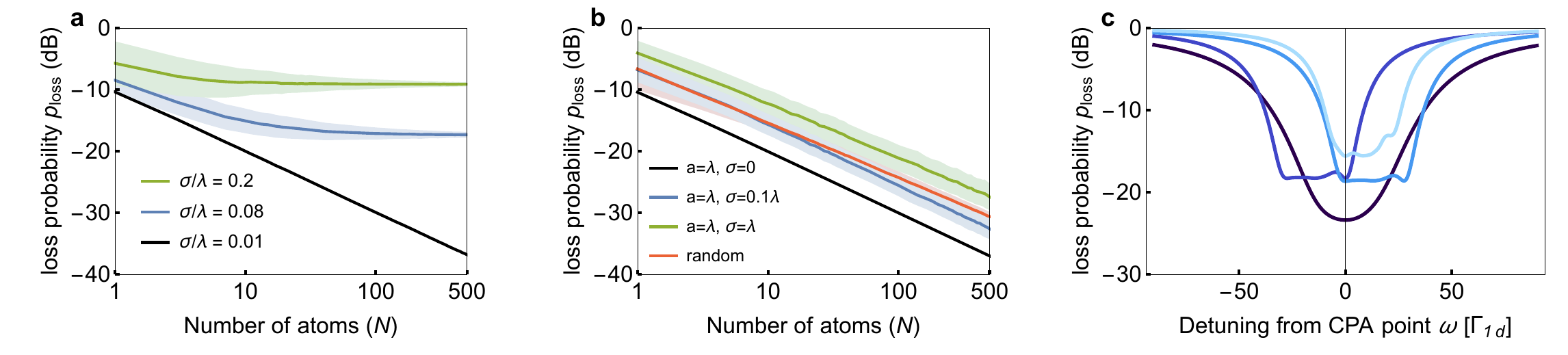}
  \caption{
  	\textbf{Photon detection in mirror geometry.}
  	\textbf{a} Photon loss probability ($p_{\mathrm{loss}}=1-\eta$) on resonance as a function of atom number $N$ for an array in the atomic mirror configuration (lattice spacing $a=\lambda$) in the presence of weak (black), intermediate (blue) and strong (green) spatial disorder (averaged over a normal distribution with standard deviation $\sigma\equiv\sqrt{\langle\delta x^2\rangle}=\{0.01, 0.08, 0.2\}\lambda$, where $\lambda$ is the wavelength of light).
  	The standard deviation of each curve is shown as lightly shaded area, averaged over 2000 realizations.
  	\textbf{b} Same system as \textbf{a}, but with engineered dissipation set to the average expected largest dissipation rate $\Gamma_{s}=\Im[\langle \mu\rangle_{\sigma,N}]$,
  	which mitigates the saturation in detection efficiency.
  	In red we show the result for a completely random array, but with fixed (characterizable) disorder.
  	\textbf{c} Photon loss as a function of frequency calculated for a completely disordered array of $N=200$ atoms when tuned to the first, second, third, and fourth eigenvalue (from dark to light blue).
  	Purcell factor for all plots is $P=\Gamma_{g}/\Gamma_\mathrm{free}=10$.
  }
  \label{fig:mirror}
\end{figure*}

One can verify explicitly that $\matr S^{-1}$ is the inverse scattering matrix
(here, $\matr \Gamma'$ is a matrix, for generality)
\begin{equation}
  	\matr S^{-1}\matr S=
  	\matr1-\matr L\dagg\matr M(\omega)\left[ \matr L\matr L\dagg-i(\matr H_{\mathrm{eff}}-\matr H_{\mathrm{eff}}\dagg) \right]\matr M\dagg(-\omega)\matr L=\matr 1,
  	\label{eq:FDT}
\end{equation}
where
\begin{equation}
  \matr M(\omega)\equiv(-i\omega+i\matr H_{\mathrm{eff}}+\matr\Gamma'/2)^{-1}.
  \label{eq:M_matrix}
\end{equation}
The square brackets in \cref{eq:FDT} vanish, which can be checked explicitly for the two examples below.
It can also be shown to hold generically, since if $\matr\Gamma'=0$, the scattering matrix is unitary $\matr S^{-1}=\matr S\dagg$,
which is only true if the term in square brackets vanishes.

Physically, this can be interpreted as the fluctuation-dissipation theorem, as the anti-Hermitian part of $\matr H_{\mathrm{eff}}$ specifies the damping, whereas $\matr L$ captures how strongly the modes are coupled to the input noise operator.

\section{Destructive photon counting}\label{sec:destructive}
\subsection{Mirror geometry}\label{sec:mirror}
We now turn to the specific model that best illustrates these ideas, namely an array of atoms coupled to a waveguide terminated by a mirror as sketched in \cref{fig:sketch}a.
In this geometry, the non-Hermitian Hamiltonian induced by integrating out the waveguide photons reads
\begin{equation}
  	\matr H_{\mathrm{eff},1,mn}=-i\frac{\Gamma_{g}}{4}
  	\left[e^{ik_0|x_m-x_n|}-e^{ik_0(x_m+x_n)}\right].
  \label{eq:mirror_Hamiltonian}
\end{equation}
where $\Gamma_{g}$ is single-atom decay rate for the transition $e\to g$ into the wave\-guide,
$x_n$ the position of the $n$th atom, 
and $k_0$ is the wavevector of the emitted light (wavelength $\lambda=2\pi/k_0$).
The coupling of the atoms via the waveguide contains both a term due to photons travelling directly in between them,
accumulating a phase $k_0|x_m-x_n|$,
and one mediated by photons being reflected from the mirror, which incurs a minus sign and a phase $k_0(x_m+x_n)$.
Since there is only one input and output field (\emph{cf.} \cref{fig:sketch}a), the matrix $\matr L$ is now a vector $L_n=\sqrt{\Gamma_{g}}\sin(k_0x_n)$.

It is instructive to see an example of how perfect absorption manifests in this setup. 
Placing the atoms in the atomic mirror configuration at positions $x_n=(1/4+n)\lambda$ (due to the infinite-range interactions, the lattice need not have unity filling),
the photonic field only couples to the symmetric collective atomic excitation $B=\sum_n b_n/\sqrt{N}$, which also is an eigenmode of the atomic array.
All other modes are dark and do not participate in the dynamics. 
In terms of this collective mode, the governing equations reduce to the input-output equations for a one-sided cavity~\cite{Collett1984} with internal dissipation
\begin{subequations}
  \begin{align}
  	&\dot B(t)=
  	-\frac{\Gamma_{\mathrm{tot}}}{2}B(t)+\sqrt{N\Gamma_{g}}a_{\mathrm{in}}(t),
  	\label{eq:AMC_langevin}\\
  	&a_{\mathrm{out}}(t)=a_{\mathrm{in}}(t)-\sqrt{N\Gamma_{g}}B(t),
  	\label{eq:AMC_io_equation}
  \end{align}
  \label{eq:AMC}
\end{subequations}
where we have introduced the total decay rate $\Gamma_{\mathrm{tot}}=N\Gamma_{g}+\Gamma'$.
As in our discussion above, $\Gamma'=\Gamma_{\mathrm{free}}+\Gamma_{s}$ comprises both free-space decay and the decay from $e\to s$, which is assumed to be tunable.
Solving Eqs~(\ref{eq:AMC}a, b) in frequency space, we calculate the number of photons in the output field 
\begin{equation}
  \langle a_{\mathrm{out}}\dagg(\omega)a_{\mathrm{out}}(\omega)\rangle
  =\left|1-\frac{N\Gamma_{g}}{\Gamma_{\mathrm{tot}}/2-i\omega}\right|^2\langle a_{\mathrm{in}}\dagg(\omega)a_{\mathrm{in}}(\omega)\rangle .
  \label{eq:output_photons}
\end{equation}
If decay is tuned such that $\Gamma_{\mathrm{tot}}=N\Gamma_g+\Gamma'=2N\Gamma_{g}$,
there is perfect absorption on resonance ($p_{\mathrm{abs}}=1$), with a bandwidth of $2N\Gamma_{g}$, 
which in this single-mode picture is equivalent to impedance matching or critical coupling.
In order to match the collective decay, the engineered decay $\Gamma_s\propto N$, such that as the atom number is increased, the detection efficiency $\eta=p_{\mathrm{abs}}\Gamma_{s}/(\Gamma_{s}+\Gamma_{\mathrm{free}})$ can become arbitrarily close to 1.

In \cref{fig:mirror}a we include spatial disorder
and show how the photon loss on resonance $p_{\mathrm{loss}}\equiv 1-\eta$ scales with atom number.
Clearly, while the setup works very well for low spatial disorder ($\sigma/\lambda<1\%$), it suffers significantly from disorder.
In the following we show how this is mitigated.
Note that in \cref{fig:mirror} and indeed all plots in this paper we choose the Purcell factor $P=10$, which close to the current state-of-the-art~\cite{Nayak2019}.
However, the collective enhancement of the atom-waveguide coupling means that the primary effect of having a lower Purcell factor is that more atoms have to be employed and therefore read out.

In the presence of disorder, the energies and decay rates of the eigenmodes of the atomic array are shifted,
and many collective atomic modes couple to the input field.
Thus, the picture presented above breaks down.
Yet, as we have shown following \cref{eq:generic_scattering_matrix}, full absorption can be attained generically, independent of disorder, by tuning the engineered dissipation $\Gamma_{s}$ to one of the eigenmodes.
However, the eigenmodes in the presence of disorder are not known \emph{a priori}, so this approach appears infeasible.
Surprisingly, one can still vastly improve over naively setting $\Gamma_{s}=N\Gamma_{g}$ as we did in \cref{fig:mirror}a,
if the standard deviation $\sigma\equiv\sqrt{\langle \delta x^2\rangle}$ of atomic positions is known.
For a given $N,\sigma$, one can then calculate the average largest eigenvalue $\langle \mu\rangle_{N,\sigma}$ and tune the engineered dissipation to its imaginary part
$\Gamma_{s}=\Im[\langle \mu\rangle_{\sigma,N}]$.
Note that $\Re\langle \mu\rangle_{\sigma,N}=0$ if every configuration is as likely as its reflection.
This restores the favourable scaling of detection efficiency with $N$, as illustrated by the blue ($\sigma=0.1\lambda$) and green ($\sigma=\lambda$) curves in \cref{fig:mirror}b.
Most strikingly, this works even in the presence of disorder equal to the lattice spacing (green), which is essentially equivalent to a fully random configuration.
The reason it works lies in the fact that the largest eigenvalue and thus the absorption bandwidth grows faster than the fluctuations of the largest eigenvalue around its mean.

If the disorder is fixed as a result of fabrication, such as in solid-state implementations, one can further improve the scaling by measuring the largest decay rate.
In this case, $\Gamma_{s}$ can be tuned exactly to the largest eigenvalue.
This situation corresponds to the red curve in \cref{fig:mirror}b,
calculated for completely random configurations, where in each case $\Gamma_{s}$ was set to coincide with the imaginary part of the largest eigenvalue.
This clearly leads to a better result than without exact tuning (green).
The scaling is still worse than with low disorder (blue),
because the largest eigenvalue grows slower in completely disordered configurations compared with mostly ordered ones.

\begin{figure*}[bt]
  \centering
  \includegraphics[width=\linewidth]{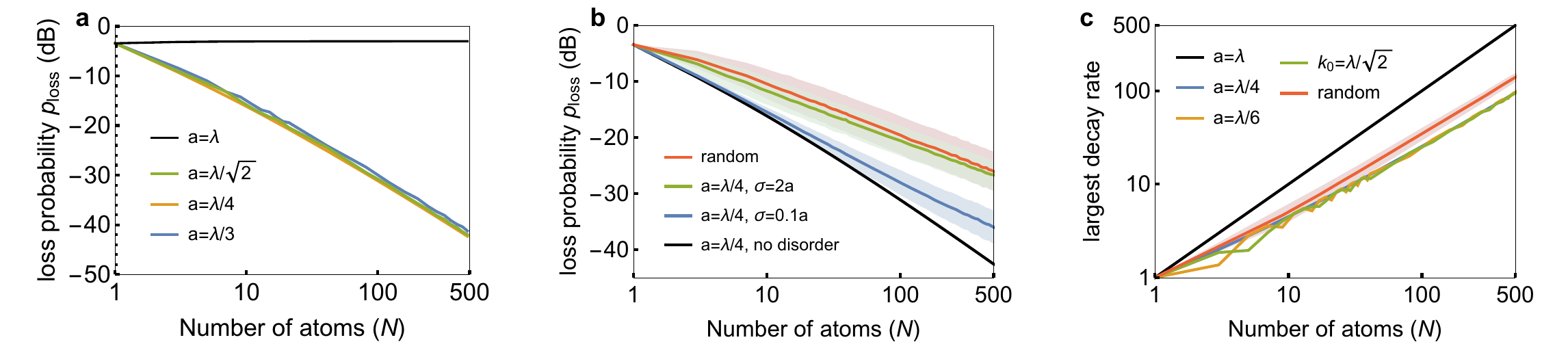}
  \caption{
  	\textbf{Photon detection with an infinite waveguide.}
  	\textbf{a} Detection probability for a disorder-free array of $N$ atoms with different spacings.
  	This graph illustrates the fact that an array in the atomic-mirror configuration ($a=\lambda$) cannot serve as photon counter,
  	whereas all other generic spacings work similarly well, a fact that can be understood analytically (\emph{cf.} \cref{sec:infinite}).
  	\textbf{b} Probability for an undetected photon on resonance for an atomic array coupled to an infinite waveguide as a function of atom number.
  	The blue line denotes the limit of a perfectly ordered array with spacing $a=\lambda/4$ (or $5\lambda/4$ etc).
  	Purcell factor is $P=10$, the average for the other cases was performed over 2500 disorder realizations (standard deviation shown as lightly coloured area).
  	\textbf{c} Scaling of largest eigenvalue in ordered arrays with varying spacing compared to a fully random one (red).
  	While the atomic mirror configuration ($a=\lambda$) is clearly different, it is a fine-tuned exception, with all other generic arrays (ordered or disordered) behaving remarkably similar.
  	The robustness of our scheme relies to a large degree on this universal eigenvalue scaling.
}
  \label{fig:infinite}
\end{figure*}

So far, we have just discussed absorption and detection on resonance. 
Equally important is the detection bandwidth, given by the engineered decay rate.
Since also the detection efficiency $\eta$ depends on the ratio between engineered dissipation and total decay rate,
best detection is achieved when tuning to the most dissipative eigenmode (\emph{cf.} \cref{fig:mirror}c).
This is our choice for all plots.

\subsection{Infinite waveguide}\label{sec:infinite}
Let us now turn to an atomic array coupled to an infinite waveguide, which has the simpler effective Hamiltonian 
\begin{equation}
 \matr H_{\mathrm{eff},1,mn}=-i\Gamma_{g}\exp(ik_0|x_m-x_n|), 
  \label{eq:infinite_Heff}
\end{equation}
since there is only one path for a photon to travel from one atom to the next.
As illustrated in \cref{fig:sketch}b, there are now two input and two output modes, a right-moving one ($+$) and a left-moving one ($-$)
[\emph{cf.} \cref{eq:generic_langevin_equations}].
The atomic lowering operators couple to the input operators via the $N\times2$ matrix $\matr L_{n,\nu}=\sqrt{\Gamma_{g}}\exp(ik_0\nu x_n)$, where $\nu\in\{\pm1\}$ labels right- and left-propagating modes.
The scattering matrix reads
\begin{equation}
  \matr S(\omega) = \mat{1&0\\0&1} -  \mat{\vec L_-\matr M(\omega)\vec L_+ &\vec L_- \matr M(\omega)\vec L_-\\
  \vec L_+\matr M(\omega)\vec L_+ & \vec L_+\matr M(\omega)\vec L_-},
  \label{eq:scattering_matrix}
\end{equation}
where $\matr M$ is defined as above [\cref{eq:M_matrix}].
Since $\matr M$ is symmetric, transmission of right- and left-moving waves (the diagonal elements of $\matr S$) are equal. 

One can show that for atoms arranged in a periodic array,
parity symmetry implies that reflection amplitude of right- and left-moving photons differs only by a phase. 
This is because $\matr J\matr M\matr J=\matr M$ under the action of the exchange matrix $\matr J$, which consists of ones on the anti-diagonal and otherwise zeros,
and $\matr J\vec L_+=\exp(i\phi)\vec L_-$ for some $\phi$.
In this case, we can parameterize the scattering matrix as 
\begin{equation}
  \matr S(\omega) = \mat{A(\omega) & B(\omega)e^{2i\phi} \\ B(\omega) & A(\omega)},
\end{equation}
with eigenvalues $\mu=A\pm B\exp(i\phi)$.
In this system, coherent perfect absorption (one eigenvalue is zero) is equivalent to $B=\exp(i\theta)A$ and $\exp(i\phi+i\theta)=\pm1$. 
In the end, full absorption of a uni-directional wavepacket may only be attained if $\matr S=0$.
Thus, parity symmetry implies that perfect absorption may only occur if the scattering matrix is zero, corresponding to an exceptional point. 
Interestingly, as the atom number $N\to\infty$, the scattering matrix $\matr S\to0$ for all arrays \emph{except} the atomic-mirror configuration. 
In the latter, the scattering matrix can be shown to reduce to
(for full absorption $\Gamma_{s}=N\Gamma_{g}$)
\begin{equation}
  \matr S_{\mathrm{AMC}}(\omega=0) = \frac12\mat{1 & -1 \\ -1 & 1},
  \label{eq:infinite_S_AMC}
\end{equation}
which gives perfect absorption for wavepackets that are symmetric superpositions of left- and right-propagating modes, but not for wavepackets incident from one direction, an effect seen before~\cite{Romero2009,Fan2013}.
This can be overcome through atomic lenses~\cite{Li2015a}, a mirror as above~\cite{Peropadre2011}, or by using any other lattice spacing~\cite{Romero2009}.
The latter can be deduced by estimating the magnitude of the elements of the scattering matrix through $|t|^N\sim\exp(-N^{1-\alpha})$,
having assumed that the largest eigenvalue scales as $N^\alpha\Gamma_{g}$.
Only in the atomic mirror configuration is $\alpha=1$, otherwise $\alpha<1$.

We numerically check the behaviour for a range of other spacings as well as fully disordered arrays and find similar behaviour as long as the atomic positions differ sufficiently from the atomic mirror configuration.
In \cref{fig:infinite}b,c we demonstrate this for a selection of array spacings (without including disorder).
We choose $k_0=\pi/(2\lambda)$ for no particular reason, except that this appears to be a good choice.
Including disorder, and employing the same technique of tuning to the average largest eigenvalue, we find similar scaling behaviour as in the mirror geometry, illustrated in \cref{fig:infinite}a.
The upshot is that arbitrary detection efficiencies can again be attained by increasing atom number, independent of disorder.

\subsection{Chiral atom--waveguide coupling}\label{sec:chiral}
Interestingly, the recently demonstrated platforms for chiral atom-waveguide coupling~\cite{Lodahl2017} are another architecture in which robust photon detection may be achieved.
Such a coupling is realized in a range of situation, for example when the light field is strongly confined \cite{Luxmoore2013,Junge2013,Shomroni2014,Sollner2015},
when giant atoms are tuned to give a chiral coupling~\cite{Kockum2018}, or in topological systems~\cite{Barik2018,Barik2019}.

By design, (almost) no backscattering occurs in these systems, there are no collective effects,
and the spacing of the atoms is immaterial, making the analysis straightforward.
If an atom coupled to right- and left-moving modes at different rates, transmission and reflection on resonance are captured by~\cite{Lodahl2017}
\begin{equation}
  \beta_\pm=\frac{\gamma_\pm}{\gamma_++\gamma_-+\Gamma'},\quad t_\pm=1-2\beta_\pm,\quad r_\pm=-2\sqrt{\beta_+\beta_-}.
  \label{eq:beta_factor}
\end{equation}
As before, $\Gamma'=\Gamma_{s}+\Gamma_{\mathrm{free}}$.
This allows us to calculate the per-atom absorption probability (for the $+$-mode)
\begin{equation}
  A_+\equiv 1-|t_+|^2-|r_+|^2 = 4\beta_+\left( 1-\beta_+\frac{\gamma_-}{\gamma_+} \right).
  \label{eq:absorption}
\end{equation}
The probability that the photon is dissipated in the right channel is $\Gamma_s/\Gamma'$ as before.
In the limit of many atoms, the transmission probability vanishes, as all photons are either reflected or absorbed.
The detection efficiency is therefore the ratio of absorbed photons times the probability they are dissipated to $\ket s$,
which to first order in $\gamma_-/\gamma_+$ is given by
\begin{equation}
  \eta_{\mathrm{chiral}}
  =\left( 1-\frac{\gamma_-}{\gamma_++\Gamma'} \right)\frac{\Gamma_{s}}{\Gamma'}.
  \label{eq:chiral_p}
\end{equation}

Note that even for moderate $\gamma_-/\gamma_+$, the second term can be reduced arbitrarily by increasing $\Gamma_{s}$, with the caveat that a larger number of emitters is needed before complete extinction is attained. 
In the absence of backscattering, this scheme is intrinsically robust against disorder. 
On top of that, the detection bandwidth depends only on the bandwidth of chirality and thus is---at least in principle---independent of $\Gamma_{s}$.
This comes again with the caveat that photons far detuned from resonance on a scale of $\Gamma_{s}$ can only be absorbed with a large number of emitters.

\section{Nondestructive photon counting}\label{sec:QND}
\subsection{Outline}
In what we have discussed so far, we have disregarded the photons emitted via the engineered decay. 
After the state of the atomic array is measured, the only information obtained concerns the number of photons in the pulse.
Since for each photon absorbed, one is emitted via the engineered channel, the question is pertinent whether a situation can arise in which the emitted photonic state coincides with the input state.
This requires the outgoing photons to be disentangled from the atoms, save for in the collective number basis, and that the photonic state is not distorted in any other way.
If these conditions are fulfilled, the setup realizes a quantum nondemolition (QND) photon-number measurement.

It is immediately obvious that if the photons are emitted into free space and thus scattered in all directions, the outgoing photonic state is a) useless, b) still entangled with the atoms, and c) distributed over many different modes.
In principle, this is mitigated to some degree if the photons are emitted back into the waveguide.
However, if each atom were to emit independently, the probability of any one photon getting lost is $1/(1+P)$, where $P$ is the Purcell factor,
which severely limits the fidelity of the scattered wavepacket for realistic Purcell factors $P$.
Furthermore, a photon has a different phase depending on the atom it is emitted from, causing residual entanglement of the photons with the atomic state, except in the atomic-mirror configuration.
Thus, QND detection requires the mirror geometry (\emph{cf.} \cref{fig:sketch}d), which we study in detail below~\footnote{We note that even in the mirror geometry the phase of the emitted photon depends on the atom is emitted from if its frequency differs from the frequency of absorbed photons.
This effect is negligible if $\delta \lambda/\lambda\ll \lambda/L$.}.

It turns out that the same trick that can make absorption robust against free-space decay and disorder---collective decay---can also be used to protect the re-emission into the waveguide.
This is achieved if the atoms are in a superposition of $\ket g$ and $\ket s$ instead of all in $\ket g$.
As we demonstrate mathematically below, this yields collective enhancement of the atom-waveguide coupling for both decay channels.
This mode of operation therefore holds one further big advantage over the non-QND operation, namely that the bandwidth of the overall detector is not limited to the single-atom bandwidth.
If the final readout of the atomic state is to give information about the number of photons,
the ground state superposition must initially posses a known number of atoms in $\ket g$.
These requirements are fulfilled by \emph{Dicke states}, fully symmetric states with a definite number of excitations
\begin{equation}
  \ket{N-m,m,0} \equiv \sqrt{\frac{(N-m)!}{N!\, m!}}[S_{sg}]^m\ket G,
  \label{eq:dicke_states}
\end{equation}
where $S_{sg}\equiv \sum_{i=1}^N\sigma_{sg}^{(i)}$ is a collective spin operator, and $\ket G$ is the state in which all atoms are in $\ket g$.

\begin{figure*}[tb]
  \centering
  \includegraphics[width=\linewidth]{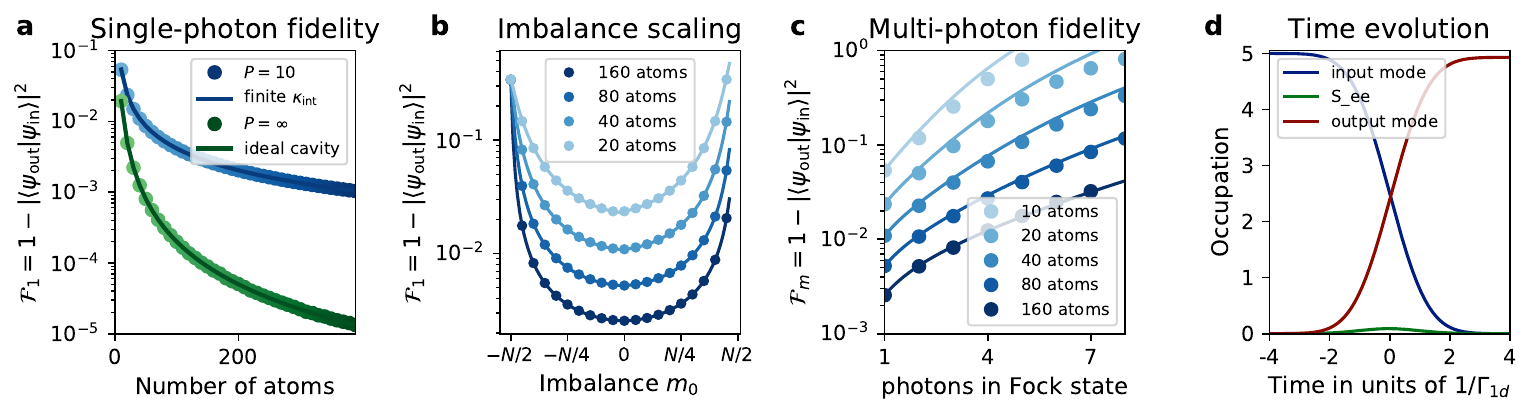}
  \caption{\textbf{QND number-resolving detection.}
  \textbf{a} Overlap $\mathcal F_1$ of the reflected wavepacket with the incoming wavepacket when a single photon in a Gaussian mode of width $\Gamma_{g}$ is scattered from a perfect array initially in state $\ket{\psi_0}=\ket{N/2,N/2,0}$,
  obtained from a numerical simulation (dots) for Purcell factors of $P=10$ (blue) and $P=\infty$ (green, no free-space decay).
  Free-space decay clearly reduces the fidelity, as is expected, but does not affect the scaling with the number of atoms, as the coherently enhanced coupling grows faster than the incoherent decay. 
  The fidelity is captured very well by the transmission fidelity of an impedance-matched two-port cavity in which each port has decay rate $\kappa=N\Gamma_{g}/2$ [black line, see \cref{eq:F1}].
  In this effective description, free-space decay corresponds to internal decay of the cavity as outlined following \cref{eq:F1}.
  \textbf{b}
  The fidelity as a function of the initial imbalance $m_0=m-N/2$, again comparing numerics with the analytical calculation for a two-sided cavity [\cref{eq:F1}].
  Here and in the following we always take $P=10$.
  \textbf{c} 
  Numerical calculation of the fidelity $\mathcal F_m$ of the QND measurement like in \textbf{a}, but now for several photons.
  The dots are numerically calculated, the solid lines is a simple estimate, given in \cref{eq:fidelity} below.
  \textbf{d} A time trace of a simulation of 5 photons in the same Gaussian mode of width $\Gamma_{g}$ scattering off 40 atoms. This illustrates that transmission is good even with modest atom numbers,
  and that the probability for an atom to be in the excited state remains small throughout, implying that free-space decay is suppressed.
  The numerical method we use is detailed in \cref{app:numerics}.
}
  \label{fig:QND}
\end{figure*}

\subsection{QND measurement with Dicke states}
We now aim to describe how a number-resolving QND measurement works in practice.
We first focus on analytical results and approximations to motivate and illustrate the idea, and then corroborate the results with numerics.
In the atomic-mirror configuration ($a=\lambda$) the coupling via a semi-infite waveguide \eqref{eq:mirror_Hamiltonian} simplifies to $\matr H_{\mathrm{eff},1,mn}=-i\Gamma_g/2$.
We also assume that the $e\to s$ transition couples to either the same waveguide at a slightly different frequency or to another waveguide field, such that
$\matr H_{\mathrm{eff},2,mn}=-i\Gamma_s/2$.
In this case, the Langevin equations \cref{eq:langevin} reduces to a single equation of motion for the collective spin operator
\begin{equation}
  \begin{aligned}
  	&\dot S_{ge}= (S_{gg}-S_{ee})\left(\sqrt{\Gamma_g}a_{\mathrm{in},1}-\frac{\Gamma_g}{2}S_{ge}\right)\\
  	&+S_{gs}\left(\sqrt{\Gamma_s}a_{\mathrm{in},2}-\frac{\Gamma_s}{2}S_{se}\right)
  	-\frac{\Gamma_{\mathrm{free}}}{2}S_{ge},
  \end{aligned}
  \label{eq:QND_eom}
\end{equation}
and another one with $g\leftrightarrow s$ and $1\leftrightarrow2$,
where $S_{\alpha\beta}=\sum_i^N\sigma^{(i)}_{\alpha\beta}$ are collective spin operators.
The input-output equations in this case read
\begin{equation}
  a_{\mathrm{out},1} = a_{\mathrm{in},1} - \sqrt{\Gamma_g}S_{ge},\quad a_{\mathrm{out},2} = a_{\mathrm{in},2}-\sqrt{\Gamma_s}S_{se}.
  \label{eq:qnd_io}
\end{equation}

Note that neglecting the input fields corresponding to free-space decay in \cref{eq:QND_eom} is an approximation, as they take the system out of subspace of Dicke states.
The reason is that when a photon decays into free-space, it destroys the coherence, which leaves the system in a mixed state.
Therefore, \cref{eq:QND_eom} fails to account properly for the dynamics after the first photon has been dissipated to free space.
The reason we can still use it to calculate the fidelity of QND measurements of Fock states is that once a photon is lost, the fidelity immediately drops to zero and remains zero, such that the subsequent time-evolution of the system is immaterial.

To understand how impedance matching can be attained here, consider the system being in the symmetric state with $m_e$ excitations
$\ket\psi=\ket{N/2-m_0-m_{\mathrm{e}},N/2+m_0,m_{\mathrm{e}}}$,
where we have defined the imbalance $m_0=m-N/2$.
Acting on this generic symmetric state,
\begin{equation}
  S_{gs}S_{se}\ket\psi = (S_{ss}+1)S_{ge}\ket\psi.
  \label{eq:symmetric_simplification}
\end{equation}
Since this is true independent of $m_0$ and $m_{\mathrm{e}}$, it is an operator identity, but only in the subspace of symmetric states.
Another identity can be obtained by exchanging $s\leftrightarrow g$.
This allows us to rewrite the equation of motion in the simplified form
\begin{equation}
  \begin{aligned}
  	\dot S_{ge}= -&\left[(S_{gg}-S_{ee})\Gamma_g + (S_{ss}+1)\Gamma_s-\Gamma_{\mathrm{free}}\right]\frac{S_{ge}}{2}
  	+\xi_{\mathrm{in}},
  \end{aligned}
  \label{eq:QND_eom2}
\end{equation}
where $\xi_{\mathrm{in}}$ is the same input noise as in \cref{eq:QND_eom}.
In symmetric states with sufficiently large population in both ground states,
we can replace the operators approximately by their expectation value in the initial state,
$S_{gg}-S_{ee}\approx N/2-m_0$ and $S_{ss}\approx N/2+m_0$.
Thus we conclude that 
\begin{equation}
  \Gamma_g(N/2-m_0) = \Gamma_s(N/2+m_0+1)
  \label{eq:impedance_matching_condition}
\end{equation}
ensures impedance matching, such that the decay rate of an excitation via the first and via the second channel are equal up to $\mathcal O(1/N)$. 
Under this condition, the equation of motion~\eqref{eq:QND_eom2} on resonance can be solved approximately 
$S_{ge}\approx a_{\mathrm{in},1}/\sqrt{\Gamma_g}$.
On the one hand, this implies that $a_{\mathrm{out},1}$ is independent of $a_{\mathrm{in},1}$, since the whole signal is absorbed.
On the other hand, $a_{\mathrm{out},2}\approx-S_{sg}a_{\mathrm{in},1}\sqrt{2\Gamma_s/\Gamma_gN^2}$.
Given this, $\langle  a_{\mathrm{out},2}\dagg a_{\mathrm{out},2}\rangle\approx \langle a_{\mathrm{in},1}\dagg a_{\mathrm{in},1}\rangle $,
\emph{i.e.}, every photon incident on port 1 is transmitted to port 2.
The presence of $S_{sg}$ in this expression indicates that for each incoming photon, an atom is transferred from $g$ to $s$. 
This still holds for the second, third, \ldots, $n^{\mathrm{th}}$ photon, but with an error of order $\mathcal O(n/N)$,
which is the same as for the non-QND detector, up to constants of $\mathcal O(1)$.

To make this intuition quantitative, we follow a simple argument.
First of all, let us denote the overlap of the output with the input wavepacket in case of the scattering of a single photon in a given state as $\mathcal F_1=|\langle\psi_{\mathrm{out}}|\psi_{\mathrm{in}}\rangle|^2$.
For any finite-bandwidth wavepacket this differs from unity. 
For example, a single photon with a Gaussian wavefunction of width $\tau$ is transmitted by an
impedance-matched cavity with outcoupling rate $\kappa$  and internal decay rate $\kappa_{\mathrm{int}}$ with fidelity
\begin{equation}
  \mathcal F_1 = \frac{\kappa}{\kappa+\kappa_{\mathrm{int}}}
  \left\{1-\sqrt{2\pi}e^{2\kappa^2\tau^2}\kappa\tau[1-\erf(\sqrt{2}\kappa\tau)]\right\},
  \label{eq:F1}
\end{equation}
where $\erf(x)=(2/\sqrt{\pi})\int_{0}^x\exp(-y^2)dy$ is the error function.
To make contact with our simulations, we thus set $\kappa=\Gamma_g(N/2-m_0)$ and the internal cavity decay $\kappa_{\mathrm{int}}=\Gamma_{\mathrm{free}}$.
In this limit of few excitations, it is therefore valid to think of free-space decay as limiting the fidelity by introducing a branching ratio between decay back into the waveguide and decay into free space, which is captured explicitly by the pre\-factor in \cref{eq:F1}.

Given some single-photon fidelity $\mathcal F_1$, a linear device would transmit $m_{\mathrm{p}}$ photons in the same mode with a fidelity of $\mathcal F_1^{m_{\mathrm{p}}}$.
However, for each atom that transitions from $\ket g$ to $\ket s$, the single-photon collective decay rate from $e\to g$ is reduced by $\Gamma_g$, whereas the collective decay rate from $e\to s$ is increased by $\Gamma_s$, which leads to imperfect impedance matching.
Using \cref{eq:output_photons} to estimate the resulting additional reflection,
we find the probability for absorbing the $k^{\mathrm{th}}$ photon is reduced by $1-2k^2/3N^2$. 
Thus, the probability that $m_{\mathrm{p}}$ photons are absorbed is reduced by a factor of $1-2(2m_{\mathrm{p}}^3-3m_{\mathrm{p}}^2+m_{\mathrm{p}})/3N^2$ to third order in $1/N$.
Combining this with the single-photon fidelity $\mathcal F_1$, we can estimate the $m_{\mathrm{p}}$-photon QND fidelity as
\begin{equation}
  \mathcal F_{m_{\mathrm{p}}} = 1-\left(1-\frac{4m_{\mathrm{p}}^3-6m_{\mathrm{p}}^2+2m_{\mathrm{p}}}{3N^2}\right)\mathcal F_1^{m_{\mathrm{p}}}.
  \label{eq:fidelity}
\end{equation}
In the following, we compare these predictions with numerical simulation and find they agree well.

\subsection{Numerical simulation}
In order to verify these conclusions numerically, we study the scattering of a multi-photon Fock states with a Gaussian wavepacket of width $\Gamma_{g}$ from an array of atoms in the atomic mirror configuration in the mirror geometry using a recently proposed technique~\cite{Kiilerich2019}.
We present details of the simulation in \cref{app:numerics}.

In \cref{fig:QND}a we show the fidelity for a single photon scattering off an array starting from the initial state $\ket{N/2,N/2,0}$. 
Clearly, \cref{eq:F1} is a good approximation to the transmission fidelity of the atomic array.
This is still true for any other symmetric starting state, as we illustrate through \cref{fig:QND}b,
which shows the fidelity of single-photon scattering when starting with an initial state $\ket{N/2-m_0,N/2+m_0,0}$, with the imbalance $m_0$ ranging from $-N/2$ to $N/2$.
We assume there is a maximally achievable decay rate $\Gamma_{\mathrm{max}}\geq\Gamma_g,\Gamma_s$ and consequently lower either $\Gamma_g$ or $\Gamma_s$ to fulfil the above condition~\eqref{eq:impedance_matching_condition}.
Together, these results show that single-photon transmission is captured very well by the above equations.

Turning to multi-photon scattering, we simulate several photons in the same Gaussian wavepacket scattering off the atomic array and again calculate the overlap of the output wavepacket with the input wavepacket.
The results are shown in \cref{fig:QND}c.
We find that our simple argument captures the fidelity well, and that our proposal can in principle reach very high fidelities for modest atom numbers.
Finally, in \cref{fig:QND}d we show an example of the time evolution of the system.

\subsection{Dicke state preparation}\label{sec:preparation}
Following similar arguments as in the other sections, the QND detector is robust against spatial disorder.
However, the suppression of free-space decay crucially relies on the preparation of a Dicke state between the two ground states. 
One way to obtain such a state is to start with all atoms in the ground state $\ket g$, apply a $\pi/2$-pulse on the ground states $\left\{ g,s \right\}$, and finally perform a projective measurement of $S_{gg}$ (or $S_{ss}$ or $S_{gg}-S_{ss}$).
This heralds a fully symmetric state with a binomial distribution of imbalances around zero and standard deviation $\sqrt{N/4}$.
However, such measurements are difficult.
In principle, one can apply an off-resonant probe in the waveguide and recording the phase shift of the reflected light, which is proportional to the number of atoms~\cite{Beguin2014}.
In \cref{app:prep} we briefly analyze this kind of measurement and find it is fundamentally limited to atom numbers of the order of the Purcell factor $N\lesssim P$.
It has been proposed to produce atomic states by manipulating the dark-state manifold~\cite{Gonzalez-Tudela2015a}, which however is limited in fidelity by $1-\mathcal F\propto N/(2\sqrt{P})$, where $N$ is the number of atoms and $P$ the Purcell factor. 
Neither method scales well to many atoms.
Thus, in the following we find a fast preparation method that imposes much less stringent requirements on the Purcell factor.

\begin{figure}[tb]
  \centering
  \includegraphics[width=\linewidth]{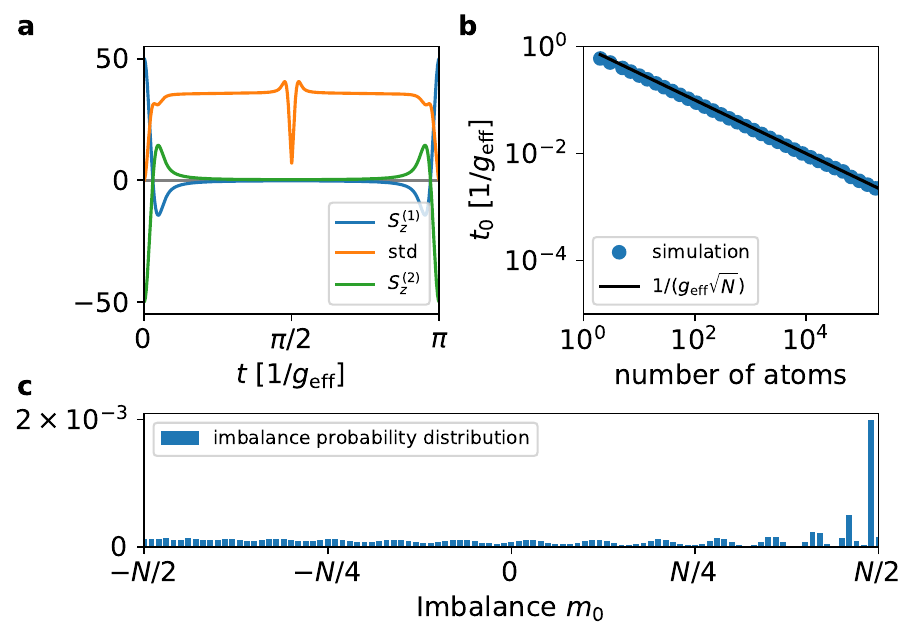}
  \caption{
  	\textbf{Dicke state preparation with \cref{eq:H_dicke}.}
  	A symmetric Dicke state can be prepared by coupling one fully excited array to one in the ground state according to Hamiltonian \cref{eq:H_dicke}.
  	\textbf{a} Time evolution of magnetization of the two arrays, $N=100$. The initial state is recovered after a period $\pi/g_{\mathrm{eff}}$.
  	After a time $\pi/2g_{\mathrm{eff}}$ the state of each array is close to $\ket{N/2,N/2,0}$ with small fluctuations in the imbalance.
  	The first equilibration happens much faster, after $g_{\mathrm{eff}}t_0=1/\sqrt{N}$,
  	which is beneficial to reduce free-space decay.
  	\textbf{b} Dependence of $t_0$ on atom number $N$.
  	\textbf{c} Imbalance distribution at time $t_0$ for $N=100$ atoms.
  }
  \label{fig:prep}
\end{figure}

We propose to produce Dicke states in a way that is in keeping with the core idea of this article: measuring atoms, not photons. 
This requires two arrays (halves of one array) that are individually addressable with external driving.
We further require that the atomic transition frequency be in the bandgap of the waveguide, which allows the atoms to be coupled coherently without dissipation.
As shown elsewhere~\cite{Douglas2015}, such a setup readily gives rise to a Hamiltonian that couples all spins
\begin{equation}
  H = \frac{2\pi g_k^2}{\Delta L}\sum_{ij}\sigma_{sg}^{(i)}\sigma_{gs}^{(j)}e^{-|x_i-x_j|/L}\simeq
  g_{\mathrm{eff}}S_{sg}S_{gs},
  \label{eq:H_dicke}
\end{equation}
where the combination of coupling to individual waveguide modes $g_k\approx\text{const}$, bound state decay length $L=\sqrt{\alpha/\Delta}$, detuning of the impurity from the band edge $\Delta=\omega_0-\omega_b$, and band curvature $\omega_k=\omega_b+\alpha k^2$ yield the
effective coupling strength $g_{\mathrm{eff}}=2\pi g_k^2/(\Delta L)$.

Importantly, when using a Raman transition ($s\to e\to g$) to couple the atoms to the waveguide modes, both $\Delta$ and $g_k$ in \cref{eq:H_dicke} are tunable.
Careful analysis (\emph{cf.}~\cref{app:band_gap_interaction}) reveals that the effective Purcell factor $P_{\mathrm{eff}}=g_{\mathrm{eff}}/\Gamma_{\mathrm{eff,free}}$
scales as the ratio of the effective waveguide density of state ($1/\sqrt{\Delta\alpha}$) and the constant free-space density of state ($\rho_0$),
\emph{viz.}, $P_{\mathrm{eff}}\propto1/(\rho_0\sqrt{\alpha\Delta})$.
The detuning from the band edge $\Delta$ can in principle be made arbitrarily small without violating the adiabatic condition or the Markov approximation (\emph{cf.}~\cref{app:band_gap_interaction}).
Note also that disorder in the positions of the atoms is not an issue here, as there is no position-dependent phase~\footnote{
  As discussed in Ref.~\cite{Douglas2015}, a dependence on position may arise, if the modes at the band edge, through which the atoms are coupled, have spatial structure, for example a finite wavevector.
  If the atoms are placed below the band edge of a waveguide, this does not happen as the lowest energy modes are extended modes at zero wavevector.
}.
Instead this scheme is likely ultimately limited by disorder in the coupling strengths or energy, as they destroy the symmetry of the effective Hamiltonian.

\begin{figure}[tb]
  \centering
  \includegraphics[width=\linewidth]{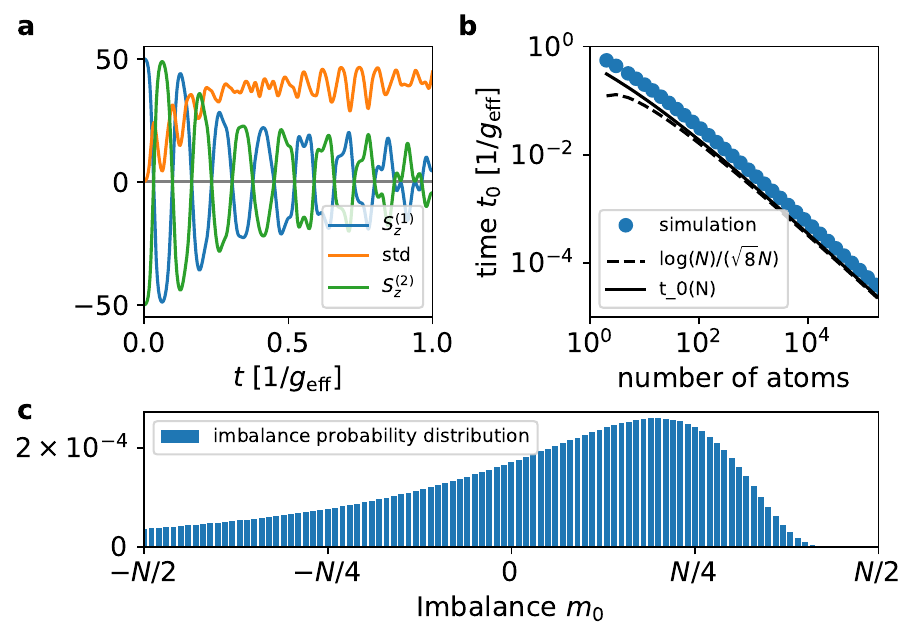}
  \caption{
  	\textbf{Fast Dicke state preparation with \cref{eq:H_dicke2}.}
  	\textbf{a} Time evolution of magnetization of the two arrays, $N=100$.
  	The time dependence can be approximated by Rabi oscillations for a few periods, which can be described in the continuum limit as outlined in \cref{app:time_scales}.
  	\textbf{b} Dependence of the zero crossing time $t_0$ on atom number $N$.
  	\textbf{c} Imbalance distribution at time $t_0$ for $N=100$ atoms.
  }
  \label{fig:prep2}
\end{figure}

The protocol to prepare an approximately half-excited state of one of the arrays is as follows.
One array is fully excited (by applying a $\pi$-pulse) and the other is left in the ground state.
The time evolution under the Hamiltonian \eqref{eq:H_dicke}, shown in \cref{fig:prep}a, transfers excitations from one chain to the other, while leaving them in their individual symmetric subspaces.
Notably, after a short time $g_{\mathrm{eff}}t_0\sim1/\sqrt{N}$,
corresponding to the first zero crossing in \cref{fig:prep}a, the average number of excitations in each array is equal.
However, this comes with the caveat that while on average the two arrays hold $N/2$ excitations each, in fact their imbalance has a very wide probability distribution, as we illustrate in \cref{fig:prep}c for $N=100$ emitters.
As we have shown above, a large imbalance does not invalidate our scheme, but it does mean that the usable atom number on average is halved.
Since this is a constant penalty, it does not change the overall scaling with atom number $N$.
Ultimately, this scheme requires $P_{\mathrm{eff}}\gg\sqrt{N}$, where $P_{\mathrm{eff}}$ is the Purcell factor enhanced through the proximity to the band edge.

There is another, intrinsically faster, way to prepare Dicke states,
if the system is governed by the Hamiltonian
\begin{equation}
  H = g_{\mathrm{eff}}[S_{sg}^{(1)}S_{gs}^{(2)}+S_{sg}^{(2)}S_{gs}^{(1)}].
  \label{eq:H_dicke2}
\end{equation}
As we detail in \cref{app:engineering_H2}, this Hamiltonian may be engineered through the interference of interactions via two different waveguides (or bands in the same waveguide), 
and constitutes a waveguide QED version of spin flip-flops recently realized in cavity QED~\cite{Davis2019}.
While certainly more challenging to implement experimentally, this Hamiltonian has the advantage of equilibrating the number of excitations in each array on an asymptotic time scale of $g_{\mathrm{eff}}t_0\sim \log(N)/(2\sqrt{2}N)$ (\emph{cf.}~\cref{app:time_scales}).
This is the fastest time that can be achieved for a given $g_{\mathrm{eff}}$, essentially saturating the time scale obtained by adding all average transition times $\sum_n^N | |S_+^{(2)}S_-^{(1)}\ket{N,n}\otimes\ket{N,N-n}| |^{-1}\sim \log(N)/2N$.
As a result, using the Hamiltonian \cref{eq:H_dicke2} reduces the requirement on the effective Purcell factor to $P_{\mathrm{eff}}\gg\log(N)$.
Another advantage of this Hamiltonian is illustrated by \cref{fig:prep2}c, which shows the probability distribution of Dicke states after a time $t_0$.
The overall probability for the state to be close to $\ket{N/2,N/2}$ is larger as with the other Hamiltonian (\emph{cf.}~\cref{fig:prep}c).

\begin{figure*}[tb]
  \centering
  \includegraphics[width=\linewidth]{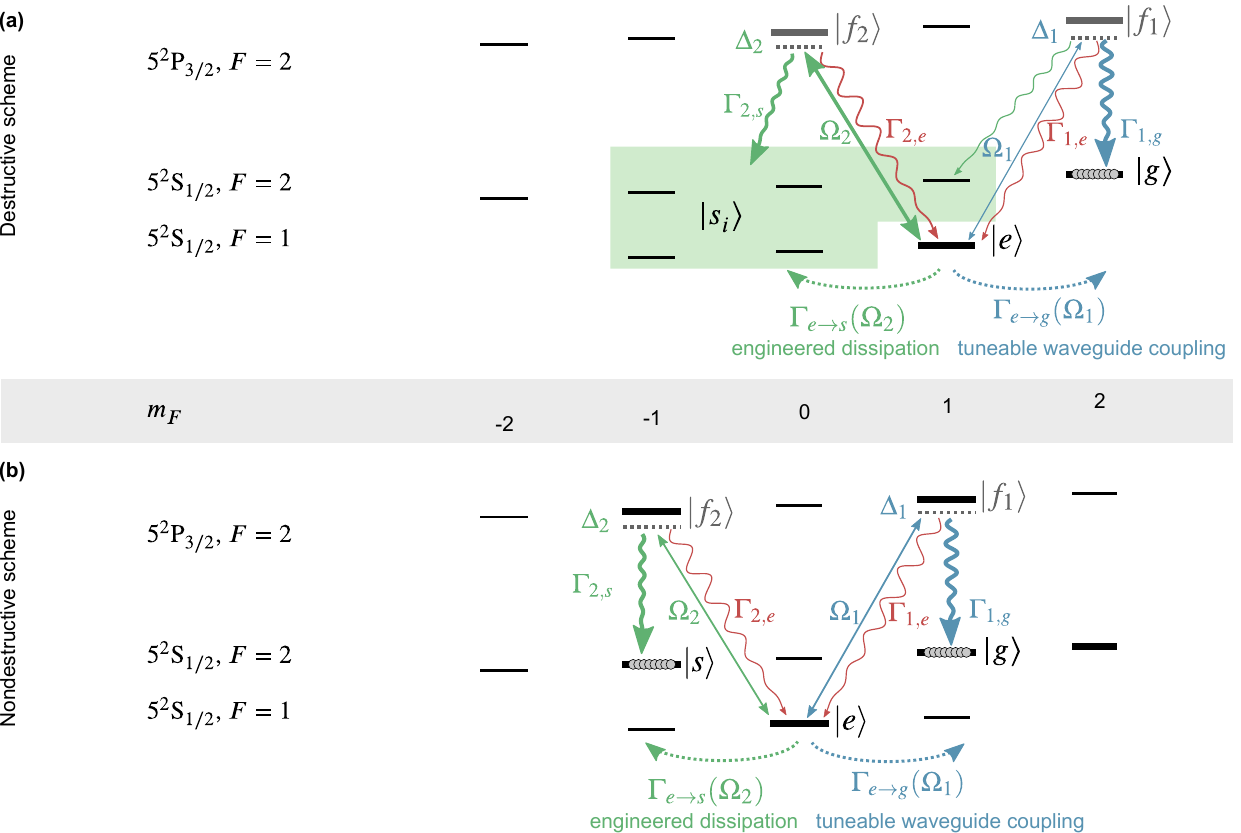}
  \caption{
  	\textbf{Double-Raman $^{87}$Rb level scheme for photon detection.}
  	Detunings $\Delta_i$ and drivings $\Omega_i$ are chosen such that the excited states can be eliminated. 
  	(a) Destructive photon detection.
  	Any of the decays into the green shaded region from $\ket{f_2}$ is intended and the total decay rate via the green channel forms $\Gamma_{\mathrm{eng}}$. 
  	Since the decay $\Gamma_{1,g}$ is superradiantly enhanced, we have $\Omega_1\ll\Omega_2$.
  	The deleterious decays shown in red are analyzed in the main text.
  	(b) Nondestructive photon detection. We have refrained from drawing in all decay channels that contribute. In comparison with the superradiantly enhanced decays $f_2\to s$ and $f_1\to g$, they scale as $\O(1/N)$ and thus can be suppressed by increasing the number of atoms.
  }
  \label{fig:level_scheme}
\end{figure*}

\section{Experimental considerations}\label{sec:experiment}
A tacit assumption in the preceding sections has been that the decay rates from $\ket e$ to $\ket g$ and/or $\ket s$ are tunable.
In circuit quantum electrodynamics, decay rates can be tuned by changing the detuning of an intermediate resonator~\cite{Pechal2014}.
In atomic systems, this can be done with Raman transitions, which we analyze in the following for the D$_2$ line of $^{87}$Rb.
For concreteness, we assume here that the waveguide efficiently couples to $\pi$ transitions.

\subsection{Engineered decay: Destructive photon measurement}

The level scheme we consider specifically is drawn in \cref{fig:level_scheme}a.
In it, we make the choice
$\ket g\equiv\ket{F=2,m_F=2}$, $\ket e\equiv\ket{1,1}$, and $\{\ket s_i\}=\{\ket{2,1},\ket{2,-1},\ket{1,0},\ket{1,-1}\}$ (green area), which are coupled via excited states (in $5^2$P$_{3/2}$)
$\ket{f_1}\equiv\ket{2,2}$, $\ket{f_2}\equiv\ket{2,0}$.
While this is just one choice among many it has the advantage that the applied lasers do not couple to any transition of the large number of atoms in the ground state $\ket g$.
Since in the destructive scheme we only need to count how many atoms have scattered photons, the final state is irrelevant, provided it is not $\ket g$.
The Raman scheme allows for large tuneability of the relative decay rate, which is required to obtain $\Gamma_{e\to s}(\Omega_2)\approx N\Gamma_{e\to g}(\Omega_1)$.

To show that the additional decays drawn in red do not spoil the scheme,
we derive the effective quantum master equation governing the double-$\Lambda$ model in their presence.
Neglecting the energy shifts due to the pumps, the dynamics are purely dissipative, given by the jump operators 
\begin{subequations}
  \begin{align}
  	\hat L_{g,\mathrm{eff}}&=\frac{\sqrt{\Gamma_{1,g}}\Omega_1}{2\Delta_1-i(\Gamma_{1,g}+\Gamma_{1,e})}\ket g\bra e,\\
  	\hat L_{s_i,\mathrm{eff}}&=\frac{\sqrt{\Gamma_{2,s_i}}\Omega_2}{2\Delta_2-i(\Gamma_{2,s_i}+\Gamma_{2,e})}\ket{s_i}\bra e,\\
  	\hat L_{ee,\mathrm{eff}}&=\sum_{i=1}^2\frac{\sqrt{\Gamma_{i,e}\Omega_i}}{2\Delta_i-i(\Gamma_{i,g_i}+\Gamma_{i,e})}
  	\ket e\bra e,
  \end{align}
\end{subequations}
where in the last expression $\Gamma_{1,g_1}=\Gamma_{1,g}$ and $\Gamma_{2,g_2}=\Gamma_{2,s}$.
We denote the (sum of the) rates corresponding to these jump operators $\Gamma_{e\to g}(\Omega_1), \Gamma_{e\to s}(\Omega_2)$, and $\Gamma_{ee,\mathrm{eff}}$, respectively.

We note that the deleterious decays $\Gamma_{i,e}$ induce dephasing described by $\hat L_{ee,\mathrm{ee}}$ that is not negligible, primarily due to the $i=2$ term, as $\Omega_2\gg\Omega_1$.
However, the dephasing removes the coherence and thus the superradiant decay to $\ket g$.
Since $\Gamma_{e\to g}/\Gamma_{e\to s}=\O(1/N)$, the corresponding excitation decays to $\ket s$, with an error $\mathcal O(1/N)$.
We conclude that all potential errors analyzed here are suppressed with increasing $N$.

\subsection{Engineered decay: Nondestructive photon measurement}

Similar considerations apply to the operation of the QND detector.
A possible choice, shown in \cref{fig:level_scheme}b contains ground states
$\ket s =\ket{2,-1}$, $\ket g =\ket{2,1}$, and excited states $\ket{f_1}=\ket{2,1}$ and $\ket{f_2}=\ket{2,-1}$.
Assuming we are able to prepare a Dicke state $|N_g=m,N_s=N-m,N_e=0\rangle$, we require the two collective decay rates to be similar [\emph{cf.}~\cref{eq:impedance_matching_condition}].

Unlike the destructive scheme, now all other decays are deleterious.
However, both decays $ e\to s$ and $e\to g$ are enhanced by a factor of $N/2$.
This implies that photon loss, either into free space or through decay into another hyperfine state scales as $1/N$. 
All deleterious decay rates can be taken together to form $\Gamma_{\mathrm{free}}$ in the calculations of \cref{sec:QND}.

\subsection{Dicke state preparation}
In order to prepare Dicke states, we propose to start with one array of the atoms in one hyperfine ground state $\ket s$ and the other array in another hyperfine ground state $\ket g$.
In this protocol, the requirements for the Purcell factor (ratio of coupling strength to free-space decay) is most stringent.
The most favourable level scheme is therefore a closed $\Lambda$-system, where $\ket g=\ket{F=2,m_F=2}$ and $\ket s=\ket{2,1}$ are hyperfine ground states, but $\ket f=\ket{3,3}$ is an excited state in $5^2$P$_{3/2}$.
In this case, $f\to g$ is a cycling transition and decays outside this subspace are suppressed.
Coherent driving between $\ket s$ and $\ket f$ can be implemented using a two-photon transition~\cite{Porras2008}.
This leaves free-space decay as error source, which has been discussed in \cref{sec:preparation}.

\subsection{Other sources of disorder}
We have neglected inhomogeneous broadening and atomic motion, which could be taken into account in the same way as positional disorder and do not modify our conclusions.
Furthermore, the relative effect of disorder decreases as the number of atoms increases~\cite{Romero2009a}.
On the other hand, fast atomic motion, analyzed in \cref{app:motion}, essentially only renormalizes the coupling strength of the atoms to the waveguide (thereby decreasing the Purcell factor), but is not a fundamental obstruction.

In typical experiments today, the atom number might fluctuate in unknown ways from one experiment to the next. 
Allowing some error in $N$ is like increasing the detuning error in $\Gamma_{\mathrm{eng}}$. As long as the relative error decreases with $N$, as it should, this does not affect the scaling with atom number.

\subsection{Photon loss from coupling into the waveguide}
It is challenging to couple photonic wavepackets travelling in free space into fibre.
If this is done with $80\%$ fidelity, say, then the detector becomes already unreliable for three photons, as there is already a 50\% chance of losing one in the first stage.

However, the detectors described here are also useful for entirely waveguide-based setups, which obviate the need for coupling free-space wavepackets into a fibre.
Recent proposals show that atomic arrays make excellent sources of quantum light~\cite{Paulisch2018,Paulisch2019,Perarnau-Llobet2019} especially in view towards quantum metrology.
There also exist proposals for two-photon gates in this platform~\cite{Zheng2013}, and it has been found that such arrays coupled to waveguides support long-lived subradiant states that can be used for storage~\cite{Zhang2019,Albrecht2019} and also manipulation~\cite{Paulisch2016} of quantum states of light.

\subsection{Atom readout}\label{sec:readout}
After the photons have scattered, one needs to measure the number of atoms in one of the states, or ideally both. 
Readout of superconducting qubits is well studied due to the advent of quantum computers.
Fidelities now reach $>99\%$ in realistic devices~\cite{Jeffrey2014,Walter2017}.

For neutral atoms, a range of techniques have been developed.
Among the earliest was to employ a cycling transition between a ground state and some excited state, which in recent experiments has yielded fidelities of 98-99\%~\cite{Kwon2017,Martinez-Dorantes2017}.
A similar idea is to push one type of atoms out of the trap with a resonant drive and use a quantum gas microscope to detect occupied sites~\cite{Nelson2007}.
Instead, one could ``boil'' one type of atom away through inelastic light scattering.
This can in principle achieve extremely high fidelities (99.97\%), but is limited by atom loss and has the drawback that the lattice has to be refilled.
We note that detection of atom numbers has already been used in experiment~\cite{Goban2015,Prasad2019}.
Likewise, a strong magnetic field~\cite{Boll2016} or a state-dependent trap~\cite{Wu2019} can be used to separate the states and image them afterwards.
In these setups, imaging has been performed with 99.94\% fidelity, but background gas collisions still cause atom loss~\cite{Wu2019}.
One option to prevent this would be tweezer arrays~\cite{Covey2019}.
While such capabilities have not yet been demonstrated for atomic arrays coupled to waveguides, and scattering from the fibre might make photon collection harder, a solution could be to move the optical lattice slowly away from the fibre before performing the imaging.

After successful detection, the detector can be reset by pumping the $s\to e$ transition, such that eventually all atoms decay to $\ket g$.

\section{Conclusion}\label{sec:conclusion}
We have explored the use of arrays of quantum emitters coupled to \mbox{waveguides} for number-resolving photon detection.
Paying particular heed to experimental limitations such as disorder and free-space decay, 
we have found that both can be overcome, leaving no fundamental limitation to the achievable detection efficiency. 
Moreover, we have shown that the same platform can also be used to perform QND number-resolving photon detection.
To this end, we also propose a novel way to prepare Dicke states based on the interaction of two arrays and subsequent heralding by measuring the number of excitations in one of them.

In a nutshell, our proposal builds on four facts that together enable highly efficient detectors:
(1), few-level systems allow for strong, projective measurements of their state due to their intrinsic nonlinearity,
(2), nevertheless, sufficiently large ensembles of atoms are linear,
(3), collective decay mitigates errors due to non-idealities, and
(4), in linear systems one can always engineer dissipation to obtain complete absorption.
We hope that the ideas outlined here will mark a step towards high-fidelity number-resolving photon detectors, both destructive and nondestructive.

\begin{acknowledgments}
  D.M.\ would like to thank Adam Smith, Clara Wanjura, and Petr Zapletal for enlightening discussions.
  D.M.\ and J.I.C.\ acknowledge funding from ERC Advanced Grant QENOCOBA under the EU Horizon 2020 program (Grant Agreement No. 742102).
\end{acknowledgments}

\appendix 
\begin{widetext}
\section{Derivation of Langevin equations}\label{app:langevin_equations}
\subsection{Two-level systems coupled to one semi-infinite waveguide field}
We study quantum emitters coupled to a semi-infinite waveguide terminated at $x=0$ by a mirror.
First assuming that the waveguide is also terminated on the other side after a length $L$, 
the bath eigenmodes have the wavefunction $\phi_n(x)=\sin(k_nx)\sqrt{2/L}$, where $k_n=\pi n / L$ for all natural $n$.
Taking the length of the waveguide to infinity, we recover the Hamiltonian given in the main text [\cref{eq:generic_hamiltonian}]
\begin{equation}
  H=\int_0^\infty\frac{dk}{2\pi}\left\{(\omega_k-\omega_0)a_k\dagg a_k-g\sum_{n}\sin(kx_n)\left(a_k\dagg \sigma^{(n)}_{ge} +\text{H.c.}\right)\right\}
  \label{eq:generic_interaction}
\end{equation}

Defining the sine transform and its inverse through
\begin{equation}
  \tilde f(\nu)=2\int_0^\infty f(t)\sin(\nu t)\,dt,\qquad 
  f(t) = \frac{1}{\pi}\int_0^\infty \tilde f(\nu)\sin(\nu t)\, d\nu,
  \label{eq:sine_transform}
\end{equation}
we can write the field in the waveguide as
\begin{equation}
  \phi(x,t) = \int_0^\infty \frac{dk}{\pi}\sin(k x)\left( a_k(t) + a_k\dagg(t) \right).
\end{equation}
Defined this way, the commutation relation $[\phi(x,t),\phi(x',t)]=\delta(x-x')$ (for positive $x$ only) implies canonical commutation relations $[a_k(t),a_q\dagg(t)]=2\pi\delta(k-q)$.
In terms of complex amplitudes, we have 
\begin{equation}
  	a(x,t) = \int_0^\infty\frac{dk}{\pi}\sin(kx)a_k(t),\qquad
  	a_k(t) =2\int_0^\infty      dx      \sin(kx)a(x,t).
\end{equation}

Solving the operator equations of motion
\begin{equation}
  	\dot a_k = -i(\omega_k-\omega_0)a_k + ig\sum_n\sin(kx_n)\sigma^{(n)}_{ge},\qquad
  	\dot \sigma^{(n)}_{ge} = ig\int\frac{dk}{2\pi}\sin(kx_n)\left( \sigma_{gg}^{(n)}-\sigma_{ee}^{(n)} \right)a_k,
\end{equation}
yields
\begin{equation}
  a_k(t) = e^{-i(\omega_k-\omega_0)t}a_k(0)+ig\int_0^t d\tau\,e^{-i(\omega_k-\omega_0)(t-\tau)}\sum_m\sin(kx_m)\sigma^{(m)}_{ge}(\tau).
  \label{eq:integrated_qle}
\end{equation}
This solution can be plugged into the equation of motion for $\sigma_{ge}^{(n)}$.
We need the following integral
\begin{equation}
  \begin{aligned}
  	&\int_0^\infty\frac{d\omega}{\pi}\sin(\omega x_n/c)\sin(\omega x/c)e^{-i(\omega-\omega_0)t}\\
  	&= \frac{e^{i\omega_0t}}{-4}\left[ \delta\left( \frac{x_n+x}{c}-t \right)+\delta\left( \frac{x_n+x}{c}+t \right)-\delta\left( \frac{x_n-x}{c}-t \right)-\delta\left( \frac{x_n-x}{c}+t \right) \right]\\
  	&= \frac{e^{i\omega_0t}}{4}\left[ \delta\left( \frac{|x_n-x|}{c}-t \right) - \delta\left( \frac{x_n+x}{c}-t \right)\right],\qquad\text{if }x,x_n,t>0.
  \end{aligned}
  \label{eq:aux_integral}
\end{equation}
The equation of motion for $\sigma_{ge}^{(n)}$ becomes
\begin{equation}
  \begin{aligned}
  	\dot \sigma_{ge}^{(n)}(t)&=\left( \sigma_{gg}^{(n)}-\sigma_{ee}^{(n)} \right)\left\{\frac{ige^{i\omega_0t}}{4}\left[ a(x_n+ct,0)-a(ct-x_n,0) \right]\right.\\
  	&+\left. \frac{g^2}{8c}\sum_m\left[ e^{ik_0(x_m+x_n)}\sigma^{(m)}_{ge}\left( t-\frac{x_m+x_n}{c} \right)-e^{ik_0|x_n-x_m|}\sigma^{(m)}_{ge}\left( t-\frac{|x_m-x_n|}{c} \right) \right]\right\}.
  \end{aligned}
\end{equation}
We define the input field $a_{\mathrm{in}}(t)$ as the portion of the waveguide field that was at a position $x=ct$ at time $t=0$ and has since travelled all the way to the atoms.
Thus, $a_{\mathrm{in}}(t) =-\sqrt{c/2}e^{i\omega_0t}a(ct,0)$, where the pre-factor is fixed by the commutation relations of $a_{\mathrm{in}}$, up to an arbitrary phase.
This yields the Langevin equation
\begin{equation}
  \begin{aligned}
  	\dot \sigma_{ge}^{(n)}(t)&=\left( \sigma_{gg}^{(n)}-\sigma_{ee}^{(n)} \right)\left\{\frac{g}{4i}\sqrt{\frac{2}{c}}\left[ e^{ik_0x_n}a_{\mathrm{in}}(t-x_n/c)+e^{-ik_0x_n}a_{\mathrm{in}}(t+x_n/c) \right]\right.\\
  	&+\left. \frac{g^2}{8c}\sum_m\left[ e^{ik_0(x_m+x_n)}\sigma^{(m)}_{ge}\left( t-\frac{x_m+x_n}{c} \right)-e^{ik_0|x_n-x_m|}\sigma^{(m)}_{ge}\left( t-\frac{|x_m-x_n|}{c} \right) \right]\right\}.
  \end{aligned}
\end{equation}
As defined, $a_{\mathrm{in}}(t)$ is a slow variable, so if the dynamics of the system and the bandwidth of the input state around $\omega_0$ are slow compared 
to the time it takes for light to travel a distance $2x_n/c$, we can neglect the retardation, rendering our description Markovian.
The same applies to the atomic lowering operators.
Finally, we arrive at a time-local equation
\begin{equation}
  \dot \sigma^{(n)}_{ge}(t)=\left( \sigma_{gg}^{(n)}-\sigma_{ee}^{(n)} \right)\left\{\frac{g}{\sqrt{2c}}\sin(k_0x_n)a_{\mathrm{in}}(t)
  + \frac{g^2}{8c}\sum_m\left[ e^{ik_0(x_m+x_n)}-e^{ik_0|x_n-x_m|}\right]\sigma^{(m)}_{ge}(t)\right\}.
\end{equation}

To calculate the output field, we take the integrated equation of motion for the light field and apply a sine transform.
This is essentially the same as the right-hand side of the equation of motion for $\sigma_{ge}^{(n)}$, except evaluated at a different point in space.
Choosing this point to be $x_R+\eps$, i.e., a small distance to the right of the rightmost atom, and again neglecting retardation, we find
\begin{equation}
  \int\frac{d_k}{\pi}\sin(kx)a_k(t)=-i\sqrt{\frac{2}{c}}\sin[k_0(x_R+\eps)]a_{\mathrm{in}}(t)-\frac{ig}{4c}\sum_me^{ik_0(x_R+\eps)}2i\sin(k_0x_m)\sigma^{(m)}_{ge}(t).
\end{equation}
Further choosing $\eps$ such that $\sin[k_0(x_R+\eps)]=1$, and defining $a_{\mathrm{out}}(t)=-\sqrt{c/2}e^{i\omega_0t}a(x_R+\eps,t)$, we have
\begin{equation}
  a_{\mathrm{out}}(t) = a_{\mathrm{in}}(t)-\frac{g}{\sqrt{2c}}\sum_m\sin(k_0x_m)\sigma^{(m)}_{ge}(t).
  \label{eq:io}
\end{equation}
Finally, let us define the decay rate $\Gamma_{g}=g^2/2c$.
\begin{subequations}
  \begin{align}
  	\dot \sigma^{(n)}_{ge}(t) &= \left( \sigma_{gg}^{(n)}-\sigma_{ee}^{(n)} \right)\left\{\sqrt{\Gamma_{g}}\sin(k_0x_n)a_{\mathrm{in}}(t)-\frac{\Gamma_{g}}{4}\sum_m\left[ e^{ik_0|x_m-x_n|}-e^{ik_0(x_n+x_m)} \right]\sigma^{(m)}_{ge}(t)\right\},\\
  	a_{\mathrm{out}}(t) &= a_{\mathrm{in}}(t)-\sqrt{\Gamma_{g}}\sum_m\sin(k_0x_m)\sigma^{(m)}_{ge}(t).
  \end{align}
\end{subequations}

\subsection{Three-level systems coupled to two semi-infinite waveguide fields}
In the main text we consider (effective) three-level systems with state $\ket g, \ket e, \ket s$ coupled to two waveguide fields, which respectively couple to 
the transition $g\leftrightarrow e$ and $s\leftrightarrow e$.
Starting from the Hamiltonian
\begin{equation}
  H=\int_0^\infty\frac{dk}{2\pi}\left[(\omega_{1,k}-\omega_0)a_{1,k}\dagg a_{1,k}+(\omega_{2,k}-\omega_0)a_{2,k}\dagg a_{2,k}
  -\sum_{n}\sin(kx_n)\left(g_1a_{1,k}\dagg \sigma^{(n)}_{ge} +g_2a_{2,k}\dagg \sigma^{(n)}_{se}+\text{H.c.}\right)\right], 
  \label{eq:three_level_interaction}
\end{equation}
the derivation follows through as above, and yields
\begin{subequations}
  \begin{align}
  	\dot \sigma^{(n)}_{ge}(t) &= \left( \sigma_{gg}^{(n)}-\sigma_{ee}^{(n)} \right)
  	\left\{\sqrt{\Gamma_{g}}\sin(k_0x_n)a_{\mathrm{in},1}(t)-\frac{\Gamma_{g}}{4}\sum_m\left[ e^{ik_0|x_m-x_n|}-e^{ik_0(x_n+x_m)} \right]\sigma^{(m)}_{ge}(t)\right\},\nonumber\\
  	&\qquad\qquad+ \sigma_{gs}^{(n)}\left\{\sqrt{\Gamma_{s}}\sin(k_0x_n)a_{\mathrm{in},2}(t)-\frac{\Gamma_{s}}{4}\sum_m\left[ e^{ik_0|x_m-x_n|}-e^{ik_0(x_n+x_m)} \right]\sigma^{(m)}_{se}(t)\right\},\\
  	a_{\mathrm{out},1}(t) &= a_{\mathrm{in},1}(t)-\sqrt{\Gamma_{g}}\sum_m\sin(k_0x_m)\sigma^{(m)}_{ge}(t).
  \end{align}
\end{subequations}
and two more equations if $1\leftrightarrow2$ and $g\leftrightarrow s$ are exchanged simultaneously.
These equations reduce to \cref{eq:QND_eom} for atoms in the atomic-mirror configuration $k_0x_n=2\pi(n+1/4)$.

\subsection{Two-level and three-level systems coupled to one and two infinite waveguide fields}
The derivation for the infinite waveguide proceeds in much the same way and can be found elsewhere~\cite{Caneva2015}.
A Hamiltonian that combines both bath couplings reads
\begin{equation}
  \begin{aligned}
  	H&=\sum_{\nu=\pm}\int\frac{dk}{2\pi}\sum_\alpha(\omega_k-\omega_0)a_{k,\nu,\alpha}\dagg a_{k,\nu,\alpha}
  	-\sum_n \left[ \sqrt{\Gamma_g}e^{i(\nu k-k_{L,1})x_n}\sigma_{eg}^n a_{k,\nu,1}+\sqrt{\Gamma_s}e^{i(\nu k-k_{L,2})x_n}\sigma_{es}^n a_{k,\nu,2}+\text{H.c.} \right].
  \end{aligned}
  \label{eq:infinite_dissipation_hamiltonian}
\end{equation}
Here, $k_{L,i}$ are phases imparted by the laser. They have little effect on the detection of incoming light.
There are two waveguide fields, $\alpha=1,2$, distinguished either in frequency, polarization, or by being in a different waveguide.
As before, $\nu\in\{\pm1\}$ labels right- and left-moving modes.

From \cref{eq:infinite_dissipation_hamiltonian}, we can derive the bath equations of motion, integrate them up and Fourier transform them~\cite{Caneva2015}
\begin{equation}
  a_{\nu,1}(x,t) = e^{i\omega_0t}a_{\nu,1}(x-ct,0)+i\sqrt{\Gamma_g}\Theta[ (x-\nu x_n)/c ]e^{ik_1(x-\nu x_n)+ik_{L,1}x_n}\sigma_{ge}^n[t-(x-\nu x_n)/c].
  \label{eq:collective_photon_field}
\end{equation}
For the other field, we have to exchange $1\leftrightarrow2$ and $g\leftrightarrow s$.
As above, we next derive the atomic equations of motion
\begin{subequations}
  \begin{align}
  	\dot \sigma^n_{ge} &= \sum_{\nu=\pm}\int\frac{dk}{2\pi}(-i\sqrt{\Gamma_g})e^{i(\nu k-k_{L,1})x_n}a_{k,\nu,1}(\sigma_{ee}^n-\sigma_{gg}^n)
  	+ (i\sqrt{\Gamma_s})e^{i(\nu k-k_{L,2})x_n}a_{k,\nu,2}\sigma_{gs}^n,\\
  	\dot \sigma^n_{se} &= \sum_{\nu=\pm}\int\frac{dk}{2\pi}(-i\sqrt{\Gamma_s})e^{i(\nu k-k_{L,2})x_n}a_{k,\nu,2}(\sigma_{ee}^n-\sigma_{ss}^n)
  	+ (i\sqrt{\Gamma_g})e^{i(\nu k-k_{L,1})x_n}a_{k,\nu,1}\sigma_{sg}^n,\\
  	\dot \sigma^n_{gs} &= \sum_{\nu=\pm}\int\frac{dk}{2\pi}(-i\sqrt{\Gamma_g})e^{i(\nu k-k_{L,1})x_n}a_{k,\nu,1}\sigma_{es}^n
  	+ (i\sqrt{\Gamma_s})e^{-i(\nu k-k_{L,2})x_n}a_{k,\nu,2}\dagg\sigma_{ge}^n,
  \end{align}
\end{subequations}
and replace the photon field [\cref{eq:collective_photon_field}]
\begin{subequations}
  \begin{align}
  	\dot \sigma_{ge}^n & = \sqrt{\Gamma_g}\sum_{\nu=\pm}e^{i(k_1\nu-k_{L,1}) x_n}(\sigma_{gg}^n-\sigma_{ee}^n)a_{\mathrm{in},\nu,1}
  	- \Gamma_g\sum_me^{ik_1|x_m-x_n|-ik_{L,1}(x_m-x_n)}(\sigma_{gg}^n-\sigma_{ee}^n)\sigma_{ge}^m\nonumber\\
  	&+\sqrt{\Gamma_s}\sum_{\nu=\pm}e^{i(k_2\nu-k_{L,2}) x_n}\sigma_{gs}^na_{\mathrm{in},\nu,2}
  	-\Gamma_s\sum_me^{ik_2|x_m-x_n|-ik_{L,2}(x_m-x_n)}\sigma_{gs}^n\sigma_{se}^m,\\
  	\dot \sigma_{se}^n & = \sqrt{\Gamma_s}\sum_{\nu=\pm}e^{i(k_2\nu-k_{L,2})x_n}(\sigma_{ss}^n-\sigma_{ee}^n)a_{\mathrm{in},\nu,2}
  	- \Gamma_s\sum_me^{ik_2|x_m-x_n|-ik_{L,2}(x_m-x_n)}(\sigma_{ss}^n-\sigma_{ee}^n)\sigma_{se}^m\nonumber\\
  	&+\sqrt{\Gamma_g}\sum_{\nu=\pm}e^{i(k_1\nu-k_{L,1}) x_n}\sigma_{sg}^na_{\mathrm{in},\nu,1}
  	-\Gamma_g\sum_me^{ik_1|x_m-x_n|-ik_{L,1}(x_m-x_n)}\sigma_{sg}^n\sigma_{ge}^m,\\
  	\dot \sigma_{gs}^n & =
  	\sqrt{\Gamma_g}\sum_{\nu=\pm}e^{i(k_1\nu-k_{L,1}) x_n}(-\sigma_{es}^n)a_{\mathrm{in},\nu,1}
  	- \Gamma_g\sum_me^{ik_1|x_m-x_n|-ik_{L,1}(x_m-x_n)}(-\sigma_{es}^n)\sigma_{ge}^m\nonumber\\
  	&+\sqrt{\Gamma_s}\sum_{\nu=\pm}e^{i(k_{L,2}-k_2\nu)x_n}\sigma_{ge}^na_{\mathrm{in},\nu,2}\dagg
  	-\Gamma_s\sum_me^{-ik_2|x_m-x_n|+ik_{L,2}(x_m-x_n)}\sigma_{ge}^n\sigma_{es}^m,\\
  	a_{\mathrm{out},\nu,1} & = a_{\mathrm{in},\nu,1}-\sqrt{\Gamma_g}\sum_ne^{i(k_{L,1}-\nu k_1)x_n}\sigma_{ge}^n,\\
  	a_{\mathrm{out},\nu,2} & = a_{\mathrm{in},\nu,2}-\sqrt{\Gamma_s}\sum_ne^{i(k_{L,2}-\nu k_2)x_n}\sigma_{se}^n.
  \end{align}
\end{subequations}
Using the description above, the Langevin equations for two-level systems on an infinite waveguide read
\begin{subequations}
  \begin{align}
  	\dot \sigma^{(n)}_{ge}(t) &= \left( \sigma_{gg}^{(n)}-\sigma_{ee}^{(n)} \right)
  	\left\{\sum_{\nu\in\{\pm1\}}\sqrt{\Gamma_{g}}e^{i\nu k_0x_n}a_{\mathrm{in},\nu}(t)-\Gamma_{g}\sum_m e^{ik_0|x_m-x_n|}\sigma^{(m)}_{ge}(t)\right\},\\
  	a_{\mathrm{out},\nu}(t) &= a_{\mathrm{in},\nu}(t)-\sqrt{\Gamma_{g}}\sum_me^{-i\nu k_0x_0}\sigma^{(m)}_{ge}(t),
  \end{align}
\end{subequations}
whereas the governing equations for three-level systems are
\begin{subequations}
  \begin{align}
  	\dot \sigma^{(n)}_{ge}(t) &= \left( \sigma_{gg}^{(n)}-\sigma_{ee}^{(n)} \right)
  	\left\{\sum_{\nu\in\{\pm1\}}\sqrt{\Gamma_{g}}e^{i\nu k_0x_n}a_{\mathrm{in},\nu,1}(t)-\Gamma_{g}\sum_m e^{ik_0|x_m-x_n|}\sigma^{(m)}_{ge}(t)\right\},\nonumber\\
  	& \qquad\qquad + \sigma_{gs}^{(n)}\left\{\sum_{\nu\in\{\pm1\}}\sqrt{\Gamma_{s}}e^{i\nu k_0x_n}a_{\mathrm{in},\nu,2}(t)-
  	\Gamma_{s}\sum_m e^{ik_0|x_m-x_n|}\sigma^{(m)}_{se}(t)\right\},\\
  	a_{\mathrm{out},\nu,1}(t) &= a_{\mathrm{in},\nu,1}(t)-\sqrt{\Gamma_{g}}\sum_me^{-i\nu k_0x_0}\sigma^{(m)}_{ge}(t),
  \end{align}
\end{subequations}
and two more equations if $1\leftrightarrow2$ and $g\leftrightarrow s$ are exchanged simultaneously.

To linearize these equations, we substitute $\sigma_{gg}\to1$, $\sigma_{ee}\to0$ and $\sigma_{ge}\to b$, where $[b,b\dagg]=1$.

\end{widetext}
\section{Preparing a Dicke state with collective measurements}\label{app:prep}
There is another way to prepare a Dicke state in our setup.
First applying a $\pi/2$-pulse to all atoms, and then measuring $S_{gg}$, \emph{i.e.,}
the number of atoms in $\ket g$,
which projects the state onto a symmetric state with a definite number of excitations.
For large $N$, $m$ has variance $\sqrt{N}$ around $N/2$, such that $|N/2-m|\ll N$.

A $\pi/2$ pulse can for example be produced by driving the two transitions $g\leftrightarrow e$ and $s\leftrightarrow e$ with lasers, each detuned by a large amount $\Delta$, or by applying a microwave tone.
After the pulse, the atomic state becomes
$\ket X = \prod_i(1/\sqrt{2})(1+\sigma_{sg}^{(i)})\ket G$,
which could optionally be written as $\ket X=2^{-N/2}\sum_m \mat{N\\ m}\ket{N-m,m,0}$.
Measuring $\hat S_{gg}$ thus probabilistically returns the state $\ket{N-m,m,0}$ where $m$ follows a binomial distribution.

A known method to measure the number of atoms is by measuring the phase shift of an off-resonant probe tone~\cite{Beguin2014}.
In order to prevent excited atoms from decaying via emission of a free-space photon (essentially removing the photon from the symmetric state) or via emission of a photon in the $e\to s$ channel (thereby inducing a transition to another Dicke state), the probe has to be far detuned.
However, the further detuned the probe is, the lower is the induced phase shift, thus requiring longer averaging times, which leads to a trade-off.
The analysis below shows that this sort of measurement is limits the number of atoms to the Purcell factor.

If it is possible to turn off $\Gamma_s$, we only have to consider free-space decay. 
Using the input-output equations, we find the output field for a weak coherent probe with amplitude $\alpha_{\mathrm{in}}$ detuned by $\Delta\gg N_g\Gamma_{g}$
\begin{equation}
  \alpha_{\mathrm{out}} = \left( 1-\frac{N_g\Gamma_g/2}{N_g\Gamma_g/2-i\Delta} \right)\alpha_{\mathrm{in}}
  \simeq\left( 1+\frac{N_g\Gamma_g}{2i\Delta} \right)\alpha_{\mathrm{in}},
\end{equation}
where $N_g$ is the number of atoms in state $\ket g$. 
Thus, the per-photon phase shift is $\varphi \simeq -\Gamma_g/2\Delta$.
The number of photons required to resolve single excitations therefore is $N_{\mathrm{p}}\propto4\Delta^2/\Gamma_g^2$.
On the other hand, for a given probe amplitude $\alpha_{\mathrm{in}}$ the probability of an atom to be excited is $\langle S_{ee}\rangle \simeq \Gamma_g N_g|\alpha_{\mathrm{in}}/2\Delta|^2$.
Thus, during the time it takes to scatter $N_{\mathrm{p}}$ photons, $\Gamma_{\mathrm{free}}N_g/\Gamma_g=N_g/P$ photons are lost.
Ultimately, this way of preparing Dicke states is thus limited to atom numbers lower than the Purcell factor.
If $\Gamma_s$ cannot be set to zero, but instead is comparable to $\Gamma_g$, then this constitutes the dominant decay channel and the requirement on the Purcell factor becomes even more stringent $P>N_gN_s$.

\section{All-to-all interaction in the bandgap}\label{app:band_gap_interaction}
\subsection{Strong coupling}
Here we give details on how using a Raman transition allows strong coupling of the atoms when brought close to the bandgap of a waveguide and discuss briefly some limitations.

Following \cite{Douglas2015}, two-level systems interacting via the modes in the bandgap 
couple at a rate
  $g_{\mathrm{eff}}=2\pi g_k^2/\sqrt{\Delta \alpha}$,
where $g_k$ is the coupling to the individual waveguide modes (assumed constant), 
$\Delta$ is the effective detuning of the renormalized transition frequency from the band edge and $\alpha$ is a parameter to characterize the band curvature at the band edge, via $\omega_k=\omega_b+\alpha k^2$.
If $g_k$ is the effective coupling rate obtained from adiabatically eliminating an excited state $\ket e$ in a Raman transition, 
it takes the form $g_k=g_{k,0}(\Omega/\delta)$, where $\Omega$ is the pump Rabi frequency and $\delta$ is the detuning of the pump from the transition.
Thus, 
\begin{equation}
  g_{\mathrm{eff}} = \frac{1}{\sqrt{\Delta}} \frac{2\pi g_{k,0}^2}{\sqrt{\alpha}}\left(\frac{\Omega}{\delta}\right)^2.
  \label{eq:full_geff}
\end{equation}
In order for the Markov approximation and the adiabatic elimination to be valid, we require $\delta\gg\Omega$ and $\sqrt{\Delta\alpha}\gg g_k$, so $g_{\mathrm{eff}}$ will be slow.
Fortunately, reducing $\Omega/\delta$ also reduces free-space decay in the same way, $\Gamma_{\mathrm{free,eff}}=\Gamma_{\mathrm{free}}(\Omega/\delta)^2$.

Since the atom-atom coupling additionally scales with the effective waveguide density of states $g_{\mathrm{eff}}\propto(\alpha\Delta)^{-1/2}$,
the Purcell factor $P_{\mathrm{eff}}=g_{\mathrm{eff}}/\Gamma_{\mathrm{eff,free}}\propto(\alpha\Delta)^{-1/2}$ can be made very large, independent of the original Purcell factor,
while preserving the Markov condition $\Delta\gg g$.
As this analysis shows, physically this relies on increasing the effective waveguide density of states.
At the same time, this has the effect of making the bound state extent and therefore the decay length of the induced interaction $L$ much larger than the extent of the atomic array,
such that the all-to-all interaction on the right-hand side of \cref{eq:H_dicke} becomes a very good approximation.

\subsection{Engineering $S_+^{(1)}S_-^{(2)}+\text{H.c.}$}\label{app:engineering_H2}
In order to instead realize a Hamiltonian with the interaction $S_+^{(1)}S_-^{(2)}+\text{H.c.}$ between two atomic arrays, one needs to combine two bands of the same waveguide (or two waveguides),
and additionally make use of the spatial profile of the induced atom-atom interaction.
The full expression for the coupling induced between the atoms through a waveguide band is given through~\cite{Douglas2015}
\begin{equation}
  H=\frac{2\pi g_k^2}{\Delta L}\sum_{ij}\sigma_{sg}^{(i)}\sigma_{gs}^{(j)}E_{k_0}(x_i)E_{k_0}^*(x_j)e^{-|x_i-x_j|/L},
  \label{eq:full_coupling}
\end{equation}
which differs from \cref{eq:H_dicke} through the addition of the spatial profile $E_{k_0}(x)$ of the bound state induced by coupling an atom at position $x$, where $k_0$ is the wavevector at the bandgap. 
A simple model of a waveguide exhibits bands with a dispersion $\omega_k\sim\cos(ka)$.
As such, there are two band edges at wavevectors $k_0=0$ and $k_0=\pi$. 
In more complex models the bandgaps might occur at different values of $k_0$, but these are generically not expected to all coincide, and certainly not in between two different waveguides.
With an otherwise constant mode profile, we can approximate $E_{k_0}(x_i)\approx e^{ik_0 x_i}$.
For simplicity we will take $k_0=\pi$, but this is by no means required. 

Combining two waveguides (waveguide bands), one with positive detuning $\Delta$, the other with the opposite detuning $-\Delta$, and placing both arrays of atoms in atomic mirror configurations, but spaced by an odd half-integer-multiple of wavelengths from each other, the two waveguides induce the interaction
\begin{equation}
  \begin{aligned}
  	H_2 &= \frac{2\pi g_k^2}{\Delta L}\left[ S_{sg}S_{gs} - (S^{(1)}_{sg}-S^{(2)}_{sg})(S^{(1)}_{gs}-S^{(2)}_{gs}) \right]\\
  	&=2g_{\mathrm{eff}}\left( S^{(1)}_{sg}S^{(2)}_{gs}+\mathrm{H.c.} \right),
  \end{aligned}
  \label{eq:H2}
\end{equation}
as required.
Note that we have taken $L$ to be larger than the whole configuration, and choosing $g_{\mathrm{eff}}$ to be the same for both waveguides,
which is not unreasonable as they can be tuned with the Raman transition.
Furthermore, we would like to note that the requirement of placing the atoms in two atomic mirror configurations spaced by an odd multiple of half a wavelength is not any more stringent that producing a single atomic mirror configuration, although it certainly is experimentally more challenging than the disordered configurations we have considered in other parts of the main text.

\subsection{Interaction time scales}\label{app:time_scales}
While the full Hilbert space of two arrays of $N$ spins each is $2^{2N}$ dimensional, 
the Hamiltonian \cref{eq:H2} connects the fully symmetric state $\ket{N,0}=\ket{N}_1\otimes\ket{0}_2$, to only $N$ other states, $\{\ket{N-m,m}\}$.
As in the main text, our convention here is that $\ket{m}$ denotes a fully symmetric spins state in which $m$ out of $N$ atoms are excited, whereas $\ket{m_1,m_2}$ denotes two arrays of $N$ atoms each, with $m_1$ and $m_2$ excitations, respectively.

In this space of $N+1$ states, the Hamiltonian \cref{eq:H2} is a matrix of the form
\begin{equation}
  \mathcal H_2=\mat{0 & a_1 & 0 &\cdots \\ a_1 & 0 & a_2 & \\ 0 & a_2 & 0 &\ddots \\
  \vdots&& \ddots&\ddots},
  \label{eq:H2_matrix}
\end{equation}
where
\begin{equation}
  a_m=2g_{\mathrm{eff}}m(N-m+1).
  \label{eq:am}
\end{equation}
As an aside, we note that this is similar to the model studied in Ref.~\cite{Christandl2004},
except with the elements of the matrix squared.
The connection arises since the model with $\sqrt{a_m}$ on the diagonal maps to the $N$-excitations subspace of two coupled harmonic oscillators $H=a\dagg b+\mathrm{H.c.}$ (leading to perfect state transfer), while here we instead study two coupled spins $H_2=2g_{\mathrm{eff}}S_+^{(2)}S_-^{(1)}+\mathrm{H.c.}$

If the index $m=x$ is understood as a spatial coordinate, the Hamiltonian $H_2$ can be rewritten as
\begin{equation}
  \begin{aligned}
  H_2&= e^{i\hat p}a(\hat x)+\mathrm{H.c.}\\
  &= 2a(\hat x)+i[\hat p,a(\hat x)]-\frac{1}{2}\{ \hat p^2,a(\hat x) \}+\mathcal O(p^3),
  \end{aligned}
  \label{eq:H2_spatial}
\end{equation}
where $[\hat x,\hat p]=i$, such that $e^{i\hat p}$ generates a translation by $-1$.
In order to get some insight into the dynamics of $H_2$, we consider the classical long-wavelength limit neglecting terms $\hat p^{n}$ with $n\geq3$ and replacing $\hat x\to x,\hat p\to p$, which yields
\begin{equation}
  H_2(x,p) = 2a(x) + a'(x) - a(x)p^2=V(x)-\frac{p^2}{2m(x)},
  \label{eq:H_classical}
\end{equation}
where
\begin{equation}
  a(x) = 2g_{\mathrm{eff}}(x+1)(N-x).
  \label{eq:a(x)}
\end{equation}
This Hamiltonian gives rise to periodic trajectories.
A more physical Hamiltonian would be obtained by sending $p^2\to-p^2$ and $H\to-H$, but the dynamics are the same.

Starting at $x(t=0)=p(t=0)=0$, the conserved energy of the particle is $H(0,0)=E_0=2g_{\mathrm{eff}}(3N-1)$.
This allows us to solve for the momentum as a function of position 
\begin{equation}
  p^2(x) = 2m(x)\left[ -E_0+V(x) \right],
  \label{eq:p(x)}
\end{equation}
which can be used to calculate the period of the orbit
\begin{equation}
  \begin{aligned}
  T(E_0)&=2\int_0^{N-2}dx\left( \frac{1}{\dot x} \right)\\
  &=\int_0^{N-2}\!\!-\frac{dx}{\sqrt{(x+1)(x-N)2x(x-N+2)}}.
  \end{aligned}
  \label{eq:TE0}
\end{equation}
The upper limit of the integral is the point at which the particle turns around, which can be found from $p(x)=0$.
For large $N$, the integral may be approximated through
\begin{equation}
  \begin{aligned}
  T(E_0)&\approx-2\int_0^{N/2}dx\left[ (x+1)2xN^2 \right]^{-1/2}\\
  &=\frac{\sqrt{2}\log(1+N+\sqrt{N(2+N)})}{N}\\
  &\approx\sqrt2\log(N)/N.
  \end{aligned}
  \label{eq:T_approx}
\end{equation}

For completeness, we mention another way to arrive at this result.
Instead of \cref{eq:H2}, consider the Hamiltonian
\begin{equation}
  H_3=g_{\mathrm{eff}}\left[(a\dagg)^2b^2+\mathrm{H.c.}\right],
  \label{eq:H_squeeze}
\end{equation}
where $a,b$ are the annihilation operators for two harmonic oscillators.
In the subspace spanned by the states $\ket{2N-2m}_1\otimes\ket{2m}_2$,
where now $\{\ket{m}\}$ are now Fock states of an oscillator,
the Hamiltonian is again given by a matrix of the same form as \cref{eq:H2_matrix},
except that now
\begin{equation}
  a_m=2g_{\mathrm{eff}}\sqrt{\left(N-m+\frac12\right)(N-m+1)m\left(m-\frac12\right)}.
  \label{eq:am_new}
\end{equation}
For large $N,m$ these are essentially the same matrix elements as before,
and one can check numerically that the dynamics is very similar for large enough $N$.
In the classical limit, we replace the annihilation operators by the amplitudes of the corresponding coherent states.
The classical mean-field equations of motion read
\begin{equation}
  \frac{d}{dt}\mat{\alpha\\ \beta}=-2ig_{\mathrm{eff}}\mat{\alpha^*\beta^2\\ \beta^*\alpha^2},
  \label{eq:squeezing_eom}
\end{equation}
with initial conditions $\beta(0)=1$, $\alpha(0)=\sqrt{2N}$.

Amplitude and phase degrees of freedom can be separated by change of variables to
$\alpha=a\exp(ix)$, $\beta=b\exp(iy)$, 
which follow equations of motion
\begin{subequations}
  \begin{align}
  	\dot a &= -ab^2\sin\phi,\qquad \dot x=-b^2\cos\phi,\\
  	\dot b &= +a^2b\sin\phi,\qquad \dot y=-a^2\cos\phi,
  \end{align}
\end{subequations}
where $\phi=2x-2y$ and we have redefined time to include the factor $2g_{\mathrm{eff}}$.
Further defining the constant $K=a^2+b^2$, and the variable $r=a^2-b^2$, 
we obtain the equations of motion
\begin{subequations}
  \begin{align}
  	\dot\phi &= 2r\cos\phi ,\\
  	\dot r   &= (r^2-K^2)\sin\phi.
  \end{align}
\end{subequations}
The shape of the resulting periodic orbits can be found by integrating $dr/d\phi$.

However, since the phase $\phi$ is ill-defined in the initial state, we can choose it to be $\phi=\pi/2$. 
In this case, the system is governed only by
\begin{equation}
  \dot r = r^2-K^2,
  \label{eq:switch_eom}
\end{equation}
which is readily integrated to yield
\begin{equation}
  \left[\tanh^{-1}\left(\frac{r}{K}\right)\right]_{r(t_i)}^{r(t_f)}=-2K(t_f-t_i).
  \label{eq:swich_soln}
\end{equation}
Since initially, $r/K\simeq 1-1/N$ and $K\simeq N$,
we conclude that the natural timescale for the switch is
\begin{equation}
  T\simeq \frac{1}{N}\tanh^{-1}\left( 1-\frac{1}{N} \right)\sim\frac{\log(N/2)}{N},
  \label{eq:switch_timescale}
\end{equation}
the same result as before.
We note that if another phase $\phi$ is chosen initially, $\phi$ first rapidly evolves to a value close to $\pi/2$, after which the system evolution is very similar.

\section{Numerical simulation of multi-photon scattering}\label{app:numerics}
In this Appendix we briefly outline the theory behind our simulation of multi-photon scattering, following Ref.~\cite{Kiilerich2019}.
The key idea is to the time-evolution of the system and bath for a specific input field using the Langevin equation~\eqref{eq:QND_eom}
in combination with the input-output equations~\eqref{eq:qnd_io}.

In general, an input wavepacket can be described through one or more modes of the waveguide in which photons are created on top of the vacuum $\ket0$,
\begin{equation}
  \ket{\psi_{\mathrm{in}}}=\prod_j \int dx_j \psi_j(x_j)a\dagg(x_j)\ket{0}.
  \label{eq:input_field}
\end{equation}
As a function of time, the field travels through the waveguide, and eventually the photons interact with the atoms locally.
Instead of this spatiotemporal description, in which a quantum system couples with constant rate to the waveguide, but the wavefunction of the photons changes in space,
it has been shown that the dynamics can be equivalently captured by modulating the coupling between the system and the input mode according to the mode shape~\cite{Kiilerich2019}.

Specifically, here we assume that all $m_p$ photons of the input field are in one mode,
such that the waveguide field can be written (at $t=0$, before the photons interact with the system)
\begin{equation}
  \ket{\psi_{\mathrm{in}}}=\left[\int_{-\infty}^0 dx\, \psi_{\mathrm{in}}(x)a\dagg(x)\right]^{m_p}\ket0,
  \label{eq:waveguide_field}
\end{equation}
with
\begin{equation}
  \psi_{\mathrm{in}}(x)=(2\pi\Gamma_g^2)^{-1/2}\exp\left(-\frac{(x/c+t_0)^2}{2\Gamma_g^2}\right).
  \label{eq:wavefunction}
\end{equation}
Since this is assumed to be the wavepacket before the interaction, it has support only for negative $x$ (we assume the wavepacket is incident from the right in \cref{fig:sketch}).
After a time $t_0$, the centre of the wavepacket (travelling at speed $c$) has arrived at the system.

Replacing the spatiotemporal evolution of wavepackets in the waveguide with a time-dependent coupling constant, the input-output equations can be replaced by a quantum master equation~\cite{Kiilerich2019}
\begin{equation}
  \dot\rho=-i[H(t),\rho]+\D[L(t)]\rho,
  \label{eq:QME}
\end{equation}
where the Hamiltonian 
\begin{equation}
  \begin{aligned}
  	H(t)= \frac{i}{2}&\left[ \sqrt{\Gamma_g}g_{\mathrm{in}}^*(t)a_{i}\dagg(S_{gg}-S_{ee}) S_{ge}\right.\\
  	-&\left.\sqrt{\Gamma_s}g_{\mathrm{out}}^*(t)a_o\dagg S_{gs}S_{se}
  -\mathrm{H.c.}\right]
  \end{aligned}
  \label{eq:SLH_H}
\end{equation}
describes coupling to the input mode,
and the jump operator
\begin{equation}
  L(t) = \sqrt{\Gamma_g}S_{ge}+\sqrt{\Gamma_s}S_{se}
  +g_{\mathrm{in}}(t)a_i + g_{\mathrm{out}}(t)a_o
  \label{eq:jump}
\end{equation}
describes the dissipation induced by the coupled waveguides.
We determine the time-evolution given by \cref{eq:QME} using QuTiP~\cite{Qutip}.

Note a subtlety when comparing with the equation corresponding to \cref{eq:SLH_H} derived by Kiilerich and M{\o}lmer~\cite{Kiilerich2019},
which features an additional term $g_{\mathrm{out}}^*(t)g_{\mathrm{in}}(t)a_i\dagg a_o$. 
This term arises when calculating the scattering from one input mode to the corresponding output mode, which in our case might be $a_{\mathrm{in},1}\to a_{\mathrm{out},1}$ and can be thought of arising from the first term in the input-output equation \cref{eq:qnd_io}. 
Here, we are interested in the scattering $a_{\mathrm{in},1}\to a_{\mathrm{out},2}$, where this term is absent.

In the above expressions, the variable couplings strengths are given by
\begin{equation}
  g_{\mathrm{in}}(t) = \frac{\tilde\psi_{\mathrm{in}}(t)}{\sqrt{1-\int_0^tdt'|\psi_{\mathrm{in}}(t')|^2}}
  \label{eq:gin}
\end{equation}
and
\begin{equation}
  g_{\mathrm{out}}(t) =-\frac{\tilde\psi_{\mathrm{out}}(t)}{\sqrt{\int_0^tdt'|\psi_{\mathrm{out}}(t')|^2}}.
  \label{eq:gout}
\end{equation}
Here, the initial time to start the simulation has been chosen to be $t=0$, and since photons are assumed to travel at speed $c$, the temporal wavefunction is related to the spatial wavefunction through $\tilde\psi_{\mathrm{in}}(x)=\psi_{\mathrm{in}}(-ct)$.
For consistency, $g_{\mathrm{in}}(t)=g_{\mathrm{out}}(t)$ are taken to be zero before $t=0$, and $\psi_{\mathrm{in}}$ should be square-integrable for positive times $\int_0^\infty dt\,|\psi_{\mathrm{in}}(t)|^2=1$.
This is true for our choice \cref{eq:wavefunction} as long as $\Gamma_{g}^{-1}\ll t_0$.
Since $t_0$ is an arbitrary offset it can always be chosen to fulfil this condition.

If the system and the input field are initially in states of known excitation number, which we assume to be the case throughout, 
then it is clear that for $m_p$ photons in the input wavepacket,
we need to consider at most $m_p$ excitations in the bosonic modes $a_o$ and $a_i$ as well as the system.
Recall that we assume a symmetric Dicke state as starting point, $\ket{\psi_{\mathrm{sys}}(0)}=\ket{N/2-m_0,N/2,0}$.
As a result, the required Hilbert space has a dimension of $(m_p+1)^3$.

We note that this formalism does not specify what shape the output wavepacket $\psi_{\mathrm{out}}$ has.
Indeed, it is generically not even true that it can be described through a single mode and neither is it true in general that all photons are emitted in the second channel rather than the first.
The advantage of this formalism in our case is that the system dynamics are independent of the choice of $\psi_{\mathrm{out}}(t)$. 
Instead, if the set of output modes considered does not comprise all modes of the physical output field,
the corresponding photons are lost.
In the extreme case, when $g_{\mathrm{out}}(t)=0$, the QME \cref{eq:QME} describes the system dynamics for a given input, but does not yield any information about the output photonic state.

Here, we use this property by setting the output mode equal to the input mode $\psi_{\mathrm{out}}(t)=-\psi_{\mathrm{in}}(t)$.
This is equivalent to asking the question how many photons are transmitted without changing the shape of the wavepacket -- \emph{i.e.,} performing a QND measurement.
If the final number of excitations in mode $a_o$ is equal to the number of input photons $m_p$, all photons have been transmitted faithfully, and all photons have been dissipated via the second channel. 
This defines the fidelity $\mathcal F$, which we plot in \cref{fig:QND}.
Mathematically, it is defined as
\begin{equation}
  \mathcal F_{m_p}=
  \bra{m_p}\tr[\rho]_{\mathrm{input,sys}}\ket{m_p},
  \label{eq:fidelity_definition}
\end{equation}
where $\ket{m_p}$ is the $m_p^{\mathrm{th}}$-Fock state of the output mode $a_o$.
This is the probability that the all photons have been transmitted in the specified output mode.

\begin{figure}[ht]
  \centering
  \includegraphics[width=\linewidth]{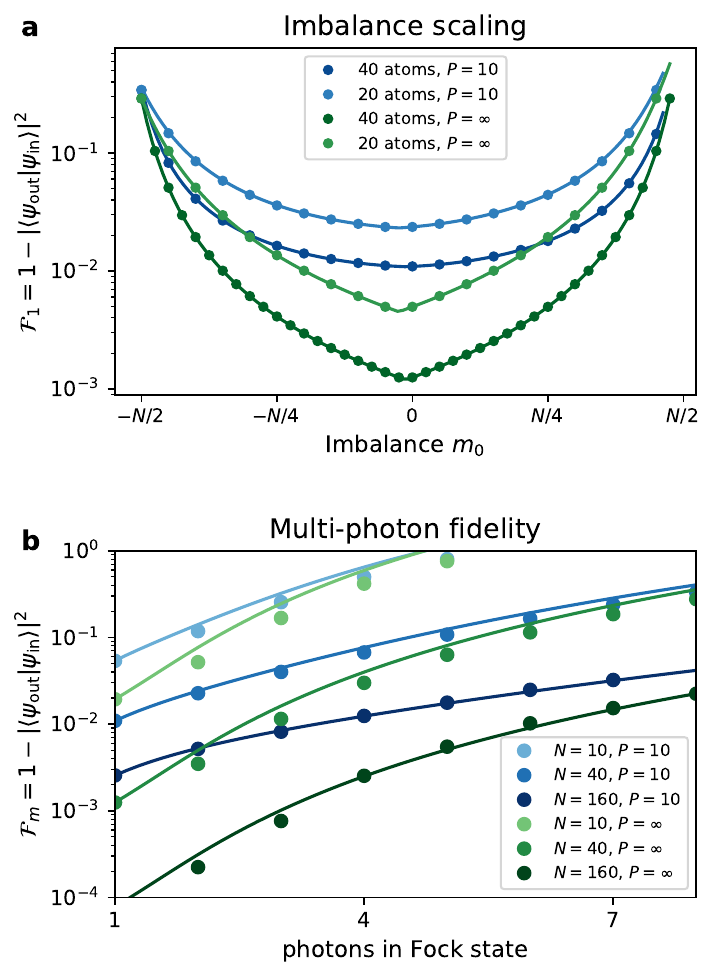}
  \caption{\textbf{Comparison between presence (blue) and absence (green) of free-space decay.}
  	Overall, the effect of free-space decay on the fidelity is clearly captured by introducing the branching ration as we have done in \cref{eq:F1}.
  	This brings down the single-photon fidelity considerably. 
  	Interestingly, though, as we consider multi-photon scattering, the effect of free-space decay diminishes relative to the effect of the non-linearities introduced.
  	Naturally, it never aids the fidelity, but in the end, the fidelity retains its favourable scaling with atom number, most clearly illustrated in \cref{fig:QND}a.
  }
  \label{fig:QND_purcell}
\end{figure}

\section{Effect of finite Purcell factor on performance of QND detector}
In this Appendix, we offer a side-to-side comparison of the fidelity with a finite Purcell factor, as shown in \cref{fig:QND} in the main text, and infinite Purcell factor. 
This allows to discern which features come from free-space decay, and which from nonlinearities and finite bandwidth. 
It furthermore offers further verification of the predicted analytical approximate fidelity \cref{eq:F1,eq:fidelity} through simulations.
The results are shown in \cref{fig:QND_purcell} and are commented on the the caption.

\section{Fast thermal motion of atoms}\label{app:motion}
In the main text we have only considered static disorder, which is valid for slowly moving atoms.
If the thermal motion of atoms is fast, some of the effect of disorder will be averaged out.
This can be modelled by instead averaging the atomic coupling over a distribution of atomic positions~\cite{Porras2008}.
Assuming a Gaussian distributions of positions around the atomic mirror configuration, captured by the random variable $y_m$
\begin{equation}
  x_m = \frac{\pi}{k_0}\left(\frac{1}{2}+2m\right)+y_m,
  \label{eq:position_fluctuations}
\end{equation}
we can calculate the off-diagonal coupling
\begin{equation}
  \begin{aligned}
  \bar g_{mn}&=-\frac{\Gamma_{g}}{8\pi\sigma^2}\iint dy_mdy_ne^{-(y_m^2+y_n^2)/2\sigma^2}\\
  &\qquad\times\left[ e^{ik_0|y_m-y_n|}+e^{ik_0(y_m+y_n)} \right].
  \end{aligned}
  \label{eq:coupling_average}
\end{equation}
If $m=n$, there should only be one integral, giving $\bar g_{nn}=-[1+\exp(-k_0^2\sigma^2)]\Gamma_{g}/(4g)$.
In the case $m\neq n$, we can straightforwardly evaluate the second term, which yields $-\Gamma_{g}e^{-k_0^2\sigma^2}/4$ overall. 
For the first term, we first shift $y_m\to y_m+y_n$, in which case the $y_n$ integral becomes a straightforward Gaussian giving a factor of $\sqrt{\pi\sigma^2}e^{y_m^2/4\sigma^2}$.
The leftover integral reads
\begin{equation}
  \begin{aligned}
  &-\frac{\Gamma_{g}}{8\sqrt{\pi\sigma^2}}\int dy_me^{-y_m^2/4\sigma^2}e^{ik_0|y_m|}\\
  &=-\frac{\Gamma_{g}}{4}e^{-k_0^2\sigma^2}\left[ 1+i\mathrm{erfi}(k_0\sigma) \right].
  \end{aligned}
  \label{eq:leftover_integral}
\end{equation}
Taken together, we get
\begin{equation}
  \bar g_{mn}=-\frac{\Gamma_{g}}{2}\left[ e^{-k_0^2\sigma^2}+\frac{i}{\sqrt{\pi}}F(k_0\sigma) \right],
  \label{eq:averaged_coupling}
\end{equation}
where $F(x)$ is the purely real Dawson integral. It is peaked at $x=1$, is odd, and obeys $|F(x)|<0.6$, and $F(x)\to0^\pm$ as $x\to\pm\infty$. 
\Cref{eq:averaged_coupling} predicts that the coupling decreases exponentially in $(k_0\sigma)^2$.
For low to moderate $k_0\sigma$, the effect of fast thermal motion can simply be accounted for by re-scaling the couplings,
without affecting the conclusions in the main text.

\bibliography{library}{}
\end{document}